\documentclass[a4paper,onecolumn,accepted=2024-01-12]{quantumarticle}
\pdfoutput=1

\usepackage[utf8]{inputenc}
\usepackage[toc,page]{appendix}
\usepackage{amsfonts}
\usepackage[dvips]{graphicx}
\usepackage{color}
\usepackage{amsmath}
\usepackage{graphicx}
\usepackage{amssymb}
\usepackage{hyperref}
\usepackage{color}
\usepackage{mathrsfs}
\usepackage{isomath}
\usepackage{amsthm}
\usepackage{epstopdf}
\usepackage{txfonts}
\usepackage{dsfont}
\usepackage{ulem}
\usepackage{physics}
\usepackage{tikz}
\allowdisplaybreaks[4]

\usepackage{bbold}
\usepackage{mathtools}

\newcommand{\ignore}[1]{} 

\newcommand{\hil}{\mathcal{H}}
\newcommand{\obsset}[1]{\boldsymbol{#1}}

\newcommand{\kb}[2]{|#1\rangle\langle #2|} 
\renewcommand{\tr}{\mathrm{tr}} 

\newcommand*{\circled}[1]{\lower.7ex\hbox{\tikz\draw (0pt, 0pt)%
		circle (.5em) node {\makebox[1em][c]{\small #1}};}}

\newcommand{\be}{\begin{equation}}
\newcommand{\ee}{\end{equation}}
\newcommand{\eea}{\end{eqnarray}}
\newcommand{\bea}{\begin{eqnarray}}
\renewcommand{\var}[1]{\ensuremath{(\Delta #1)^2}}
\newcommand{\av}[1]{\ensuremath{\langle{#1} \rangle}}

\newcommand{\Cov}{{\rm Cov}}
\renewcommand{\vec}[1]{\boldsymbol{#1}}

\newcommand{\diag}{{\rm diag}}

\newcommand{\id}{\mathbb{1}}

\newcommand{\idmap}{{\rm id}} 

\newcommand{\B}{K}

\newcommand{\Covn}[1]{{\rm Cov}_{\varrho_{#1}}(\obsset{g}_{#1})}

\newcommand{\C}{X_\varrho}

\newcommand{\DiffMat}{\mathfrak \Delta}

\newtheorem{theorem}{Theorem}

\newtheorem{corollary}{Corollary}

\newtheorem{observation}{Observation}

\newtheorem{lemma}{Lemma}

\usepackage{cleveref}
\crefname{equation}{Eq.}{Eqs.}
\crefname{observation}{Obs.}{Obs.}
\crefname{corollary}{Corollary}{Corollaries}
\crefname{lemma}{Lemma}{Lemmata}
\crefname{proof}{Proof}{Proofs}
\creflabelformat{proof}{#2proof#3}
\crefname{remark}{Remark}{Remarks}
\crefname{prop}{Proposition}{Propositions}

\begin{document}

\title{Bounding entanglement dimensionality from the covariance matrix}

\author{Shuheng Liu}
\orcid{0000-0001-7130-1888}
\affiliation{State Key Laboratory for Mesoscopic Physics, School of Physics, Frontiers Science Center for Nano-optoelectronics, \& Collaborative Innovation Center of Quantum Matter, Peking University, Beijing 100871, China}
\affiliation{Vienna Center for Quantum Science and Technology, Atominstitut, TU Wien,  1020 Vienna, Austria}
\affiliation{Institute for Quantum Optics and Quantum Information (IQOQI), Austrian Academy of Sciences, 1090 Vienna, Austria}

\author{Matteo Fadel}
\orcid{0000-0003-3653-0030}
\affiliation{Department of Physics, ETH Zürich, 8093 Zürich, Switzerland}

\author{Qiongyi He}
\orcid{0000-0002-2408-4320}
\email{qiongyihe@pku.edu.cn}
\affiliation{State Key Laboratory for Mesoscopic Physics, School of Physics, Frontiers Science Center for Nano-optoelectronics, \& Collaborative Innovation Center of Quantum Matter, Peking University, Beijing 100871, China}
\affiliation{Collaborative Innovation Center of Extreme Optics, Shanxi University, Taiyuan, Shanxi 030006, China}
\affiliation{Hefei National Laboratory, Hefei 230088, China}

\author{Marcus Huber}
\orcid{0000-0003-1985-4623}
\email{marcus.huber@univie.ac.at}
\affiliation{Vienna Center for Quantum Science and Technology, Atominstitut, TU Wien,  1020 Vienna, Austria}
\affiliation{Institute for Quantum Optics and Quantum Information (IQOQI), Austrian Academy of Sciences, 1090 Vienna, Austria}

\author{Giuseppe Vitagliano}
\orcid{0000-0002-5563-3222}
\email{poissongdb@gmail.com}
\affiliation{Vienna Center for Quantum Science and Technology, Atominstitut, TU Wien,  1020 Vienna, Austria}
\affiliation{Institute for Quantum Optics and Quantum Information (IQOQI), Austrian Academy of Sciences, 1090 Vienna, Austria}

\begin{abstract}
High-dimensional entanglement has been identified as an important resource in quantum information processing, and also as a main obstacle for simulating quantum systems. Its certification is often difficult, and most widely used methods for experiments are based on fidelity measurements with respect to highly entangled states. Here, instead, we consider covariances of collective observables, as in the well-known Covariance Matrix Criterion (CMC)~\cite{guhnecova} and present a generalization of the CMC for determining the Schmidt number of a bipartite system. This is potentially particularly advantageous in many-body systems, such as cold atoms, where the set of practical measurements is very limited and only variances of collective operators can typically be estimated. To show the practical relevance of our results, we derive simpler Schmidt-number criteria that require similar information as the fidelity-based witnesses, yet can detect a wider set of states. We also consider paradigmatic criteria based on spin covariances, which would be very helpful for experimental detection of high-dimensional entanglement in cold atom systems. We conclude by discussing the applicability of our results to a multiparticle ensemble and some open questions for future work.
\end{abstract}

\maketitle

\section{Introduction}

Entanglement has been always seen as a crucial property of quantum physics~\cite{schrod35}. Currently it is regarded also as an important resource for quantum information tasks and, at the same time, as a main obstacle for classical simulations of quantum systems~\cite{HorodeckiEntanglementReview2009,GuheneToth09,FriisNatPhys19,Frerot2023}. In fact, a lot of research has been devoted recently to the problem of distinguishing separable from entangled states, and even quantifying entanglement as a resource~\cite{PlenioVirmani07}.
Relevant entanglement measures can be given in many ways, including entropies of marginals, optimal decompositions of a state in terms of separable states, robustness to noise, distance to the set of separable states, and so on~\cite{PlenioVirmani07}. 
All of these measures are in practice very hard to estimate from experimental data, typical methods are entanglement Hamiltonian learning~\cite{Kokail2021Quantum,kokail2021entanglement}, replica trick~\cite{Islam2015}, or compressed sensing methods~\cite{Gross2010Quantum}. Therefore one typically relies on 
bounding them from so-called entanglement witnesses, i.e., observables that have positive expectation values on separable states but can have negative expectation values for some entangled states.
Bounds on entanglement measures have been discussed also in terms of nonlinear witnesses, typically involving variances of collective operators~\cite{GittsovichPRA10,FadelVitagliano_2021}.
However, such bounds work well either for very low-dimensional systems in the case of entropic measures~\cite{GittsovichPRA10} or for measures related to the tolerance of the entangled state to noise~\cite{FadelVitagliano_2021,brandao05,CramerPlenioWunderlich11,marty14}.

One particularly important measure that is relevant especially in the context of classical simulation of quantum systems is the {\it Schmidt number}, or {\it entanglement dimensionality} which, loosely speaking, quantifies the dimension needed to reproduce the correlations in the quantum state: Proving that a quantum state with a certain entanglement dimensionality $r$ means that such a state cannot be simulated with a system of dimension lower than $r$. At the same time, genuine high-dimensional entanglement has been proven useful for several practical tasks, ranging from improved security for quantum cryptography~\cite{etcheverry2013quantum,HuberPawlowski13,DodaQuantum2021}, to noise resistant quantum communication \cite{ecker19,hu20} and universal quantum computation~\cite{Lanyon2009,VandenNest13}.

Like other entropic measures, such as the concurrence or the entanglement of formation, bounding this measure from experimental data is typically very hard. Usual methods rely on fidelity measurements that must be fine-tuned and, especially in many-body systems, typically inaccessible~\cite{FriisNatPhys19,Frerot2023}. 
Because of this, entanglement dimensionality quantifications have been successfully done experimentally only in certain two-photon systems, where virtually arbitrary local measurements can be performed~\cite{ecker19,hu20,Krenn6243Generation,Erker2017Quantifying,BavarescoNatPhys18,schneeloch2019quantifying,HerreraValencia2020highdimensional}.  
On the other hand, it would be highly desirable to have more powerful tools that would allow to detect the entanglement dimensionality in other quantum platforms, e.g., cold atoms, which are suitable for the simulation of other quantum systems. This raises the question on how to find bounds to the entanglement dimensionality based on measurements routinely performed in such systems, such as simple two-body correlators, or variances of simple collective observables. In fact, this is currently a topic which has been subject of investigation under various approaches~\cite{Pichler2016Measurement,euler2023detecting}.

The approach that we follow here is inspired by the various results which relate entanglement and the squeezing of collective spin variances. In particular, we first consider the general problem of looking for a generalization of the well-known Covariance Matrix Criterion (CMC) to witness the Schmidt number of a bipartite quantum state.
After deriving the general criterion, we specialize to simpler statements that are more readily applicable to quantum experiments and compare them to the existing criteria based on fidelities with respect to pure (high-dimensionally entangled) states. 
Afterwards, we also look for criteria based on the covariances of three local spin observables, which would connect our results with the spin-squeezing methods successfully applied in many experiments, especially with atomic gases~\cite{giovannettietalPRA03,Lucke2014Detecting,VitaglianoPlanar,PezzeRMP2016,vitagliano16,BaccariDepth,FadelGessnerPRA20}. 
In fact, bipartite entanglement has been recently demonstrated between spatially separated atomic ensembles~\cite{JulsgaardNATURE2001,Fadel18,KunkelEtAl2018,Lange18,VitaglianoFadeletAl2022}, a quantification of entanglement via monotones can be provided~\cite{Islam2015,FadelVitagliano_2021,CramerEtAl2013}, and even experimentally accessible bounds on distillable entanglement in cold atoms systems for quantum information applications have been proposed~\cite{Bergh2021Experimentally,bergh21}, but not yet those associated with applications of the entanglement dimensionality.

\section{Methods}

\subsection{Definition of entanglement and Schmidt number}

Let us consider a bipartite system with Hilbert space $\hil = \mathbb C^{d_a} \otimes \mathbb C^{d_b}$.
Any arbitrary bipartite pure state $\ket{\psi} \in \hil$ can be brought to the normal form under local unitaries called {\it Schmidt decomposition}: 
\be
\ket{\psi} \mapsto U_a \otimes U_b \ket{\psi} = \sum_k \sqrt{\lambda_k} \ket{k_a k_b} ,
\ee
where $\{\ket{k_n}\}$ with $n=a,b$ are orthonormal basis for the two local Hilbert spaces and $\lambda_k$ are called {\it Schmidt coefficients}. 
Such a Schmidt decomposition can be achieved essentially from the singular value decomposition of the coefficient matrix of the state $\ket{\psi}=\sum_{kl} c_{kl}\ket{kl}$.
In other words, the transformation that brings the coefficient matrix $c_{kl}$ in its singular value decomposition amounts to a local unitary transformation of $\ket{\psi}$, which in turn provides the Schmidt decomposition. 
The Schmidt decomposition also gives the eigendecomposition of the reduced density matrix of each party as 
\be
\varrho_a = \tr_b( \ketbra{\psi} ) = \sum_k \lambda_k \ketbra{k_a} , 
\ee
and similarly for $\varrho_b$.
Such a transformation cannot map a product state into a non-product state, which is called entangled, nor vice-versa. Moreover, it cannot change the amount of entanglement of the state, since all entanglement monotones are 
by definition invariant under local unitary transformations.

In fact, the number $s(\ket{\psi})$ of nonzero Schmidt coefficients of a pure state defines the entanglement monotone called Schmidt rank. 
This monotone can be extended to all bipartite density matrices via {\it convex-roof construction}, this way defining the {\it Schmidt number}~\cite{Barbara2000SchmidtNumber,Sanpera2001SchmidtNumber}:
\be\label{eq:SNdefinition}
s(\varrho):= \inf_{\mathcal D(\varrho)} \max_{\ket{\psi_i} \in \mathcal D(\varrho)} s(\ket{\psi_i}) ,
\ee
where the infimum is taken over all pure state decompositions $\mathcal D(\varrho)=\{p_i, \ket{\psi_i} \}$ of the density matrix $\varrho=\sum_i p_i \ket{\psi_i}\bra{\psi_i}$.

\subsection{Schmidt number witnesses, $r$-positive maps and matrix norms}

One common way to lower bound the Schmidt number is through entanglement witnesses, i.e., operators $W$ such that 
\be
\begin{aligned}
&\tr(W \varrho) \geq 0 \quad  \text{for all} \quad \varrho \quad \text{such that} \quad s(\varrho) \leq r \\
&\tr(W \varrho^*) < 0  \quad  \text{for some} \quad \varrho^* \quad \text{such that} \quad s(\varrho^*) > r .
\end{aligned}
\ee
The canonical Schmidt-number witnesses are fidelities with respect to target states~\cite{Krenn6243Generation,Erker2017Quantifying,BavarescoNatPhys18,FlammiaLiu11}. 
As the most common example, considering 
a target maximally entangled state 
\be
\ket{\psi_+}=\sum_{k=1}^d \tfrac 1 {\sqrt{d}} \ket{kk} , 
\ee
one obtains that the bound valid for all Schmidt-number-$r$ states is 
\be\label{eq:fidcrit}
\tr(\varrho \kb{\psi_+}{\psi_+})\leq \tfrac r d ,
\ee
that is, a common Schmidt-number witness is given by
\be\label{eq:fidelityWit}
W_{\psi_+} = \frac r d \id - \kb{\psi_+}{\psi_+} .
\ee
Note that many similar entanglement witnesses are often considered. However, there are states such that their Schmidt numbers cannot be witnessed by any such witnesses based on the fidelity to a target state. These states have been studied in \cite{WeilenmannPRL2020BeyondFidelity} and have been termed {\it unfaithful}, or more precisely, a state is $r$-unfaithful if it has Schmidt number higher than or equal to $r$, but this cannot be witnessed by any fidelity with respect to a pure state. 

Dual to entanglement witnesses, there are (linear) quantum maps $\mathcal M(\varrho)$ that are able to witness the Schmidt number. 
In particular, transformations that map quantum states into quantum states are always positive, i.e., $\mathcal M(X) \geq 0$ for all positive inputs $X$ (hence density matrices included).
However, there are also maps that act positively only on a subset of states, e.g., maps that transform separable states into other valid quantum states, and entangled states potentially into non-physical, i.e., non-positive matrices.
In particular, in a bipartite system of dimension $r \times d$ one can consider maps which are so-called {\it $r$-positive}, i.e., maps such that 
\be\label{eq:rpositivemap}
\idmap_r \otimes \mathcal M (\varrho_{ab}) \geq 0 ,
\ee
where in this case $\mathcal M$ acts on the space of $d \times d$ matrices and  $\idmap_r(\cdot)$ is the identity map, acting on the space of $r \times r$ matrices. 
In other words, a map that is positive per s\'e but acts only on one party in a bipartite quantum state, does not necessarily maintain its positivity property. 
Whenever it does so acting on one party, with the other party that is left untouched having dimension (at most) $r$, such a map is called $r$-positive. 
Maps that are $r$-positive for all integer numbers $r$ are called {\it completely positive}.

Now, let us consider a map that is positive, but not $2$-positive and let it act only on a subsystem of a bipartite quantum state as in \cref{eq:rpositivemap}. From the definition, one can readily see that a product state will always be mapped into a positive matrix, since
\be
\idmap \otimes \mathcal M \left(\varrho_a \otimes \varrho_b \right) = \idmap \otimes \mathcal M \left(\varrho_a \otimes \varrho_b \right) = \varrho_a \otimes \mathcal M(\varrho_b) \geq 0 ,
\ee
and similar statement holds for separable state due to linearity of the map.
On the other hand, because for such maps \cref{eq:rpositivemap} does not necessarily hold even when $r=2$, there might be entangled states $\varrho_2$ such that
\be
\idmap \otimes \mathcal M (\varrho_2) \ngeq 0 .
\ee

One prominent example of such a map, is the transposition. In fact, one of the most famous entanglement criteria was derived by Peres \cite{peresPPT} and states that for all separable states it must hold that
\be\label{eq:PPTcrit}
\idmap \otimes \mathcal T (\varrho) \geq 0 ,
\ee
where $\mathcal T (\cdot)$ is the transposition map, defined on a basis of matrices as
\be
\mathcal T (\kb{j}{k}) = \kb{k}{j} .
\ee

More in general, it is known that maps that are $(r-1)$-positive and act on a subsystem of a bipartite quantum state as in \cref{eq:rpositivemap} will always map a state with Schmidt number $(r-1)$ into another positive matrix \cite{Barbara2000SchmidtNumber}. 
On the other hand, if such a map is not $r$-positive, then it can happen that there is a state $\varrho_{r}$ with Schmidt number $r$ such that
\be
\idmap_a \otimes \mathcal M (\varrho_r) \ngeq 0 .
\ee
Actually, an even more precise result holds, that a state $\varrho_{ab}$ has Schmidt number smaller than or equal to $(r-1)$ {\it if and only if} it satisfies \cref{eq:rpositivemap} {\it for all} $(r-1)$-positive maps.

One of the most typical $r$-positive maps used for Schmidt-number detection is given by a generalization of the {\it reduction map}~\cite{reductionCritH,reductionCritC}, namely 
\be
\mathcal D_r (A) = \tr(A) \id - \tfrac 1 r A .
\ee
When this map is applied locally on a bipartite density matrix one gets
\be\label{eq:rRedCrit}
\idmap_a \otimes \mathcal D_r (\varrho ) = \varrho_a \otimes \id_b - \tfrac 1 r \varrho \geq 0 ,
\ee
and its positivity must hold for all states with Schmidt number smaller than or equal to $r$~\cite{Barbara2000SchmidtNumber}.
The so-called reduction criterion is obtained by applying this map with $r=1$.

The dual nature between non-completely positive maps and entanglement witnesses is given by the Choi-Jamio\l kowski isomorphism, 
which is basically a canonical mapping between the space of operators in a bipartite space and quantum maps. 
More precisely, given a map $\mathcal M(X) : \mathbb C^{d_a} \otimes \mathbb C^{d_a} \rightarrow \mathbb C^{d_b} \otimes \mathbb C^{d_b}$ one defines a bipartite operator (sometimes called Choi matrix) as
\be
W_{\mathcal M} := d_b \idmap \otimes \mathcal M \left( \kb{\psi_+}{\psi_+} \right) ,
\ee 
where the output space is $\mathbb C^{d_b} \otimes \mathbb C^{d_b}$ and $\ket{\psi_+}$ is the maximally entangled state in a bipartite Hilbert space isomorphic to $\mathbb C^{d_a} \otimes \mathbb C^{d_a}$.

Note that this map is dual to the witness with respect to the maximally entangled state, namely, the witness in \cref{eq:fidelityWit} with $\ket{\Psi_r}$ being the $r$-dimensional maximally entangled state is obtained from the map
$\idmap_a \otimes \mathcal D_r$ after applying the Choi-Jamio\l kowski isomorphism.

\subsection{The covariance matrix criterion}

Other criteria for witnessing the Schmidt number are given in terms of local unitarily invariant norms of correlation matrices~\cite{ChenPRL05,deVicentePRAr07,klockl2015characterizing,JohnstonKribs15}, or variances of a full local basis set of observables~\cite{deVicentePRAr07}.

Variances and covariances of collective observables can be read from the (symmetric) covariance matrix associated to a pair of local observables' basis $\obsset{g} = (\obsset{g}_a, \obsset{g}_b)$, where  $\obsset{g}_a=(g_1^{(a)} \otimes \id , \dots , g_{d_a^2}^{(a)} \otimes \id)$ and $\obsset{g}_b=(\id \otimes g_1^{(b)} , \dots , \id \otimes g_{d_b^2}^{(b)})$ are orthonormal basis of observables for party $a$ and $b$ respectively\footnote{Here we consider basis operators (ortho)normalized as $\tr(g_k g_l)=\delta_{kl}$.}.
As a common example, in the case of $d=2$, one local orthogonal basis (LOB) is given by $\obsset{g} = \left( \tfrac{\id}{\sqrt{2}}, \tfrac{\sigma_x}{\sqrt{2}} , \tfrac{\sigma_y}{\sqrt{2}} , \tfrac{\sigma_z}{\sqrt{2}}\right)$, where the $\sigma_k$ are the Pauli matrices. In higher dimension, one common construction for a LOB is given by $\obsset{g} = \left( \tfrac{\id}{\sqrt{d}}, \vec \sigma \right)$, where $\vec \sigma$ is a basis of the $su(d)$ algebra of the corresponding dimension. 
However, apart from those, other LOBs can be constructed, not necessarily of the form given above.
See \cref{app:LOBs} for details.
The covariance matrix calculated on a generic $\varrho$, assumes the block form
\begin{equation}\label{eq:blockCov}
\Cov_\varrho(\obsset{g}) := \Gamma_\varrho = 
\left(\begin{array}{ll}
\gamma_a & \C \\
\C^T & \gamma_b
\end{array}\right),
\end{equation}
in which the diagonals $\gamma_a:=\Covn{a}$ and $\gamma_b:=\Covn{b}$ are the symmetric covariance matrices of each party 
\be
[\Covn{n}]_{jk} = \tfrac 1 2 \av{g_j^{(n)} g_k^{(n)} + g_k^{(n)} g_j^{(n)}}_{\varrho_n} - \av{g_j^{(n)}}_{\varrho_n} \av{g_k^{(n)}}_{\varrho_n}
\ee
with $n=a,b$, and the off-diagonal blocks\footnote{In the following we will often simplify the notation and whenever clear from the context omit the dependence on the density matrix and the local bases.}
\be
\begin{aligned}
(\C)_{kl} &= \av{g_k^{(a)} \otimes g_l^{(b)}}_\varrho - \av{g_k^{(a)}}_\varrho \av{g_l^{(b)}}_\varrho 
\end{aligned}
\ee
are the cross-covariances between the two local observables vectors. 
This matrix can be brought in a block singular value decomposition with a suitable local orthogonal transformation $O \Gamma_\varrho O^T$, with $O=O_a \oplus O_b$. 
This corresponds to an orthonormal change of local bases $\obsset g \mapsto \obsset g^\prime = O \obsset g$.

It is well-known that every separable density matrix $\sigma=\sum_k p_k \varrho_a \otimes \varrho_b$ with $p_k \geq 0$ and $\sum_k p_k=1$ has to satisfy 
\be\label{eq:CMCcrit}
\Gamma_\sigma \geq \kappa_a \oplus \kappa_b ,
\ee
where $\kappa_a = \sum_k p_k \Covn{a}$ and $\kappa_b = \sum_k p_k \Covn{b}$ are positive matrices. 
In other words, if a state is separable, there must exist positive matrices $\kappa_a$ and $\kappa_b$, which are covariance matrices of single particle states, such that \cref{eq:CMCcrit} holds.
Thus, given two positive matrices $\kappa_a$ and $\kappa_b$ that lower bound $\Gamma_\varrho$ for a certain density matrix $\varrho$ one has to be able to find pure local states $\ket{\phi_a}$ and $\ket{\phi_b}$ and probabilities $p_k$ such that $\kappa_a = \sum_k p_k {\rm Cov}_{\phi_a}(\obsset{g}_{a})$ and similarly for party $b$.
If such states $\ket{\phi_a}$ and $\ket{\phi_b}$ are not found for any such lower bound $\kappa_a \oplus \kappa_b$, then $\varrho$ must be entangled.

This is called {\it covariance matrix criterion} (CMC) in the literature \cite{guhnecova,GittsovichPRA10,gittsovich08} and is known to be equivalent to all possible entanglement criteria of the form $\sum_k \var{T_k}_\varrho \geq K_a + K_b$,
where $T_k= A_k \otimes \id + \id \otimes B_k$ are collective observables, $\var{T}_\varrho=\av{T^2}_\varrho-\av{T}_\varrho^2$ is their variance and the bounds are given by $K_a = \min_{\phi} \sum_k \var{A_k}_\phi$ and similarly for $K_b$~\cite{hofman03}. 
The CMC follows from essentially two properties: (i) the {\it concavity} of the covariance matrix under mixing quantum states, i.e., the inequality $\Cov_\varrho(\obsset g) \geq \sum_k p_k \Cov_{\varrho_k}(\obsset g)$ that holds for every mixed state $\varrho=\sum_k p_k \varrho_k$ and for every observables' set; (ii) the fact that $\av{A\otimes B}_\sigma=\av{A}_\sigma \av{B}_\sigma$ holds for all pure separable states $\sigma$ and for all local observables pairs $(A,B)$, implying that $X_{\sigma}=0$.
The CMC criterion has been proven stronger than many of the other known entanglement criteria~\cite{gittsovich08}, which are the ones later generalized to detect the Schmidt number~\cite{klockl2015characterizing,ChenPRL05,deVicentePRAr07}.

Furthermore, a quantifier of entanglement was defined from the violation of the CMC with the following idea~\cite{GittsovichPRA10}. 
First, for every quantum state $\varrho$ (also entangled) one can find a real and positive $t$ such that
\be
\Gamma_\varrho - t \kappa_a \oplus \kappa_b \geq 0 .
\ee
In the worst case, the equation above will hold for $t=0$ for all states. 
Then, one can ask the question, given a state $\varrho$, what is the maximal value of $t$ for which such an equation holds? This will quantify the minimal violation of the CMC by a quantum state. 
If such a maximum value of $t$ coincides with $t=0$ it means that the state maximally violates the CMC.  
Formally, one defines the quantity
\begin{equation}
V(\varrho)=\max _{t, \kappa_A, \kappa_B}\left\{t \leq 1: \gamma(\varrho)-t \kappa_A \oplus \kappa_B \geq 0\right\} ,
\end{equation}
and from this an entanglement parameter 
\begin{equation}
\mathcal{E}(\varrho)=1-V(\varrho) ,
\end{equation}
which is such that $\mathcal{E}(\varrho) = 0$ for separable states and $\mathcal{E}(\varrho) \leq 1$ for all states. 

Then, one might think that such a parameter would also provide an entanglement monotone. 
While the latter property turns out not to be true, namely $\mathcal{E}(\varrho)$ is not an entanglement monotone, it turns out to give lower bounds on the concurrence, and potentially on other entanglement measures which are based on convex roofs of functions of the Schmidt coefficients~\cite{GittsovichPRA10}.
Roughly speaking, the reason for which $\mathcal{E}(\varrho)$ is not an entanglement monotone, is because there are local operations, called {\it filterings}, that can increase its value.
See also \cref{app:LOBs} for some more details on these filtering operations.

Nevertheless, it is natural to think that such an entanglement parameter could provide a lower bound on the entanglement dimensionality as well.
However, it is easy to observe that this is actually not the case, as the parameter $\mathcal{E}(\varrho)$ can get its maximal value, namely $\mathcal{E}(\varrho)=1$ even on states
with Schmidt number equal to two. 
See \cref{app:tparbound} for more details.

Because of this, a meaningful generalization of the CMC to witness the Schmidt number is still lacking, which is the main goal of our work, together with elucidating its potential practical applications and its comparison with the other known criteria.

\section{A CMC criterion for entanglement dimensionality} 

Here we present our main results. First a CMC criterion with a tight bound that is valid for states with a given Schmidt number $r$. 
Afterwards, we present a corollary that is more practical for applications. 
Our main theorem follows:
\begin{theorem}\label{theo1}
Let us consider a bipartite density matrix $\varrho \in \mathcal B(\hil)$ such that its Schmidt number is $s(\varrho) \leq r$ and its covariance matrix $\Gamma_{\varrho}=\Cov_{\varrho}(\obsset{g})$ for some canonical bases' pair $\obsset{g}$. 
The matrix inequality 
\be\label{eq:CMCrankr}
\Gamma_{\varrho} \geq \sum_i p_i \Gamma^{(i)}_r 
\ee
must hold for some probability distribution $\{p_i\}$ and some positive $\Gamma^{(i)}_r=\Cov_{\psi^{(i)}_r}(\obsset{g})$, 
which are covariance matrices of pure Schmidt-rank-$r$ states.

The pure states $\Gamma^{(i)}_r$ have the block form (omitting the index $i$ for simplicity)
\be
\left(\begin{array}{cc}
\kappa_a & X_{\psi_r} \\
X_{\psi_r}^T & \kappa_b
\end{array}\right) ,
\ee
with the nonzero singular values of the blocks being 
\be 
\begin{aligned}\label{eq:kappaCconds}
\vec{\epsilon}(\kappa_a) &= \vec \epsilon(\kappa_b) = (\vec{\epsilon}(D),\{\tfrac 1 2 (\lambda_k+\lambda_l)\}_{k=1,k<l}, \{\tfrac 1 2 (\lambda_k+\lambda_l)\}_{l=1,k>l} ) , \\
\vec{\epsilon}(|X_{\psi_r}|) &=(\vec{\epsilon}(D),\{\sqrt{\lambda_k \lambda_l}\}_{k=1,k<l},\{\sqrt{\lambda_k \lambda_l}\}_{l=1,k>l})
\end{aligned}
\ee 
where $\kappa_a:=\Cov_{\psi_r}(\obsset{g}_a)$ and $\kappa_b:=\Cov_{\psi_r}(\obsset{g}_b)$. Here $\lambda_k$ and $\lambda_l$ with $1\leq k ,l  \leq r$ are the Schmidt coefficients of $\ket{\psi_r}$ and $D$ is the  
matrix with elements $D_{kl}= \lambda_k \delta_{kl}-\lambda_k \lambda_l$. 
\end{theorem}

{\it Proof.---}The statement follows from two properties: (i) concavity of covariances, i.e., the fact that $\Gamma_\varrho \geq \sum_k p_k \Gamma_{\varrho_k}$ holds for density matrices decomposed as $\varrho=\sum_k p_k \varrho_k$; and (ii) the fact that a generic covariance matrix of a pure Schmidt-rank-$r$ state has the above block singular value decomposition in a canonical basis constructed from its Schmidt bases.
This last fact can be directly verified with simple algebra (see \cref{app:LOBs} and also, e.g., \cite{GittsovichPRA10}). 
With these two ingredients, the proof goes as follows. Consider a generic Schmidt-number-$r$ density matrix $\varrho=\sum_k p_k \ketbra{\psi_r^{(k)}}$ decomposed in terms of pure Schmidt-rank-$r$ states $\ketbra{\psi_r^{(k)}}$. Using concavity we have $\Gamma_\varrho \geq \sum_k p_k \Gamma_{r}^{(k)}$ with each of the $\Gamma_{r}^{(k)}$ having 
singular values as in \cref{eq:kappaCconds}.
Then, we use the fact that arbitrary changes of local bases (which include all local unitary transformations on the quantum state) correspond to local orthogonal transformations  $O_a \oplus O_b \Gamma_\varrho (O_a \oplus O_b)^T$. In particular, such transformations cannot change the eigenvalues of $\gamma_a$ and $\gamma_b$ and the singular values of $X_{\psi_r}$.
\qed

The matrix formulation of the criterion is per s\`e very hard to evaluate due to the fact that one should in principle scan essentially all possible pure Schmidt-rank-$r$-state covariance matrices $\Gamma_{r}$. 
However, it is also possible to find corollaries to our main theorem which can be actually readily applied, as soon as even just partial information about  $\Gamma_\varrho$ is available.
In particular, considering unitarily invariant norms $\| \cdot \|$, i.e., norms that are functions of the singular values, of the covariance matrix blocks we can derive conditions which are already quite powerful.

For example, we can consider the trace norm of the cross-covariance block $X_{\psi}$. 
Since it is a quantity that is invariant under local orthogonal transformations, we can calculate it for pure states directly from \cref{eq:kappaCconds} and the value will be the same for all local orthogonal bases. This way we find the following bound
\be\label{eq:trXpsibound}
\tr|X_{\psi_r}| = \sum_k (\lambda_k - \lambda_k^2) + 2 \sum_{k<l} \sqrt{\lambda_k  \lambda_l} \leq 1 - \sum_k \lambda_k^2 + (r-1) = r - (1-E_L(\ket{\psi_r})) ,
\ee
where $r$ is the Schmidt-rank of $\ket{\psi_r}$ and 
\be
E_L(\ket{\psi_r}) = 1 - \tr[(\tr_a \kb{\psi_r}{\psi_r} )^2] = 1-\sum_k \lambda_k^2
\ee
is its linear entanglement entropy. The bound follows from $2\sum_{k<l} \sqrt{\lambda_k  \lambda_l} \leq \sum_{k<l} (\lambda_k + \lambda_l) = r-1$ and is valid for all pure states with Schmidt rank $r$. 
In turn, such a bound can be employed in our CMC criterion to find a criterion valid also for mixed states with a given Schmidt number, as we prove next. 

\begin{corollary}\label{cor:1}
The following inequality
\begin{align}
\tr|\C|- (r-1) &\leq \sqrt{
[1 - \tr(\varrho^2_a)][1 -\tr(\varrho^2_b)] }, \label{eq:cor1} 
\end{align}
holds for every density matrix $\varrho$ such that $s(\varrho)\leq r$. 
\end{corollary}

{\it Proof.---}Consider the matrix $\DiffMat := \Gamma_{\varrho} - \sum_k p_k \Gamma^{(k)}_r$, which due to \cref{theo1} must be positive for all Schmidt-number-$r$ states. Positivity of a block matrix implies the inequality (see e.g., \cite{HornJohnson})
\be\label{eq:hyperb}
\| \DiffMat_a \| \cdot \| \DiffMat_b \| \geq \left\| (\DiffMat_X^T\DiffMat_X)^{1/2} \right\|^2 ,
\ee
where we have labelled the blocks $\DiffMat_{a,b,X}$ in analogy with a generic covariance matrix. The above inequality is equivalent to the following family of inequalities
\be\label{eq:tangents}
\| \DiffMat_a \| \geq - 4 
\left( t^2 \| \DiffMat_b \| - |t| \| (\DiffMat_X^T\DiffMat_X)^{1/2} \| \right) ,
\ee
where $t$ is a real parameter. Geometrically, these can be seen as the family of tangents $y \geq -4(t^2 x -|t| c)$ to the hyperbola given by the product condition $xy\geq c^2$ in \cref{eq:hyperb}.
Note in fact, that taking the infimum of \cref{eq:tangents} over all $t \in \mathbb R$ we recover \cref{eq:hyperb}.
The rest of the proof consists in taking the trace norm $\| A \| = \tr\sqrt{A^\dagger A}:=\tr|A|$ as a special case and exploiting the bound $\tr|X_{\psi_r}|\leq r- 1 + E_L(\ket{\psi_r})$ which we derived previously.
Taking the trace norm of a matrix in \cref{eq:tangents} we obtain 
\be
\tr(\DiffMat_a) + 4 t^2 \tr(\DiffMat_b) \geq 4 |t| \tr|\DiffMat_X| \geq 4 |t| (\tr|X_\varrho| - \tr|X_{\psi_r}|),
\ee
where last we substituted the expression $\DiffMat_X=X_\varrho - X_{\psi_r}$ and used the triangle inequality.
Since $\DiffMat_a$ and $\DiffMat_b$ are positive (being principal minors of $\DiffMat$) we have 
\be
\tr(\DiffMat_a)= \tr(\gamma_a) - \tr(\kappa_a) = 1-\tr(\varrho^2_a)-E_L(\ket{\psi_r})
\ee
and analogously for $\tr(\DiffMat_b)$ where $E_L(\ket{\psi_r})$ is the linear entanglement entropy of the generic (optimal) pure Schmidt rank-$r$ state $\ket{\psi_r}$. Here we used that $\tr(\gamma_\varrho) = d - \tr(\varrho^2)$ holds for a generic single particle covariance matrix.
Furthermore, from \cref{eq:trXpsibound} we have that $\tr|X_{\psi_r}|\leq r- 1 + E_L(\ket{\psi_r})$.
Substituting these relations into the inequality above we obtain
\be
1-\tr(\varrho^2_a)-E_L(\ket{\psi_r}) + 4 t^2 (1-\tr(\varrho^2_b)-E_L(\ket{\psi_r}))\geq 4 |t| (\tr|X_\varrho| - r+ 1 - E_L(\ket{\psi_r})) , 
\ee
which can be rearranged to 
\be
(1-4|t|+4t^2)\geq (1-4|t|+4t^2)(1 - E_L(\ket{\psi_r}))\geq 
\tr(\varrho^2_a)+4t^2\tr(\varrho^2_b)+4 |t| (\tr|X_\varrho| - r) . 
\ee
Thus, we have that 
\be
(1-4|t|+4t^2) - \tr(\varrho^2_a)-4t^2\tr(\varrho^2_b)-4 |t| (\tr|X_\varrho| - r) \geq 0
\ee
holds for all values of $t$. 
Minimizing the left-hand side over $t$ we get that the minimum is achieved for  $2|t|=(\tr|X_\varrho|-r+1)/(1-\tr(\varrho_b^2))$ and results in \cref{eq:cor1}. \qed

A proof similar to the above could be made by using also other unitarily-invariant norms, as soon as bounds on the norms of the covariance matrix blocks valid for all pure Schmidt-rank-$r$ states are found. 
In particular, also submatrices constructed from specific sets of operators can be considered, as we will
explicitly do in the discussion on applications of our method.
Note also that several criteria for the Schmidt number are based on unitarily-invariant norms of the cross-correlation matrix, i.e., the linear part of the cross-covariance matrix $(\mathfrak{X}_\varrho)_{kl}:=\av{g_k^{(a)} \otimes g_l^{(b)}}_\varrho$~\cite{ChenPRL05,deVicentePRAr07,klockl2015characterizing,JohnstonKribs15}.
Similar criteria can be derived from \cref{eq:cor1}, and in general from \cref{theo1}. 
A detailed investigation of these criteria has been done in a separate publication, connected with the idea of evaluating such conditions from randomized measurements~\cite{LiuRandomized}. In this sense, our condition could be used to detect the Schmidt number without the need of tuning the measurement directions or the requirement of a common reference frame between the particles. See also~\cite{Wyderka2023Probing,Imai2023Work}. In that paper, we also discuss further details about the practical implementation of the condition with randomized measurements, including, e.g., the potential limitations coming from finite statistical samples. However, such a condition can be also evaluated from more common measurements, which are essentially the same that are required for calculating, e.g., the fidelity with respect to a maximally entangled state. These kind of measurements have been implemented several times recently with success to detect high-dimensional entanglement in photonic experiments~\cite{FriisNatPhys19,ecker19,hu20,Krenn6243Generation,Erker2017Quantifying,BavarescoNatPhys18,schneeloch2019quantifying,HerreraValencia2020highdimensional,steinlechner2017distribution,Malik_2016,bulla2022}. At the same time, the requirement of measuring correlations between full local bases becomes impractical when the local dimensions become extremely high and the measurements are very limited, such as, e.g., in cold atoms. Because of that, we discuss later ideas to derive even simpler conditions from our main \cref{theo1}.

\section{Applications}

To show the usefulness of our results we are going to discuss some initial applications of our method, which might serve also as the basis for future developments.

\subsection{Comparison with other common criteria}

First of all, we look at important classes of states that can be potentially detected, as compared to other Schmidt number criteria. By far, the most widely used is given by the fidelity witness with respect to maximally entangled states.
We can observe that already \cref{eq:cor1} is strictly stronger than every such witness.

\begin{observation}\label{obs:fidcritcompMES}
Every state $\varrho$ that violates the criterion $\tr(\varrho \kb{\psi_+}{\psi_+})\leq \tfrac r d$ for some $r$, also violates \cref{eq:cor1} for the same $r$.
\end{observation}

{\it Proof.---}The Hilbert-Schmidt scalar product between a generic density matrix $\varrho$ and $\kb{\psi_+}{\psi_+}$ can be expressed as 
\be
\tr(\varrho \kb{\psi_+}{\psi_+})=\tfrac 1 d \sum_k \av{g_k^{(a)} \otimes g_k^{(b)}}_\varrho .
\ee
Then, let us consider the criterion in \cref{eq:cor1}.
Using the fact that $\tfrac{x + y} 2 \geq \sqrt{xy}$ holds for all real numbers, we find that a weaker version is 
\be
\tfrac 1 2 (2-\tr(\varrho^2_a)-\tr(\varrho^2_b))
\geq \sqrt{
[1 - \tr(\varrho^2_a)][1 -\tr(\varrho^2_b)] } \geq \tr|\C|- (r-1) , 
\ee
which can be rewritten as
\be
\tr(\varrho^2_a)+\tr(\varrho^2_b)+2\tr|X_{\varrho}| \leq 2r . 
\ee
Considering the left-hand side of this expression we have 
\be
\begin{aligned}
\tr(\varrho^2_a)+\tr(\varrho^2_b)+2\tr|X_{\varrho}| 
&\geq \sum_{k} \av{g^{(a)}_{k}}^2+ \av{g^{(b)}_{k}}^2- 2 \av{g^{(a)}_{k}}\av{g^{(b)}_{k}} + 2 \av{g^{(a)}_{k}\otimes g^{(b)}_{k}} \\
&\geq \sum_{k} 2 \av{g^{(a)}_{k}\otimes g^{(b)}_{k}} \\
&=2d \tr(\varrho \kb{\psi_+}{\psi_+})
\end{aligned}
\ee
since $\av{g^{(a)}_{k}}^2+ \av{g^{(b)}_{k}}^2- 2 \av{g^{(a)}_{k}}\av{g^{(b)}_{k}}=(\av{g^{(a)}_{k}} - \av{g^{(b)}_{k}})^2$ are positive terms. 
Thus, we have that 
\be
\tr(\varrho \kb{\psi_+}{\psi_+}) > r/d
\ee
implies that \cref{eq:cor1} is violated.
\qed

Note that the case of $r=1$ has been already studied previously. The CMC criterion is strictly stronger than the CCNR criterion\cite{gittsovich08}, therefore stronger than the fidelity criterion with respect to $\kb{\psi_+}{\psi_+}$.

We can also observe that \cref{eq:cor1} detects states that are not detected by {\it any fidelity criterion}. Concretely, we can make a statistics on a class of randomly generated density matrices, considering a mixture of the form 
$\varrho_{\rm mix}=p (U_a\otimes U_b) \ketbra{\psi_\lambda} (U_a\otimes U_b)^{\dagger}+(1-p)\ketbra{\psi_{+}^k}$, where $\ket{\psi_{+}^k}$ is the canonical $k$-dimensional maximally entangled state, $\ket{\psi_\lambda}$ is a random pure Schmidt-rank-$4$ state, the value of the mixture $p$ is randomly and uniformly selected between $0 \leq p \leq 1$ and $U_a\otimes U_b$ are random local unitaries. The goal is to detect the Schmidt number of $\varrho_{\rm mix}$ which is a mixture of a Schmidt-rank-$4$ state and a Schmidt-rank-$k$ state. Generating a sample of $10000$ random states for each $k$, we observe that \cref{cor:1} detects higher Schmidt numbers for $10.8\%,16.4\%,37.2\%,7.6\%$ of the states when $k=4,3,2,1$ respectively, while the optimal fidelity witness detects higher Schmidt numbers for only $3.4\%,3.4\%,3.6\%,3.8\%$
of the states when $k=4,3,2,1$ respectively. 

\cref{eq:cor1} can be also seen as a nonlinear improvement~\cite{GuehneLutkenhaus06,guhnemechler} over a criterion containing the singular values of the cross-correlation matrix, namely $\sum_k \epsilon_k (\mathfrak{X_\varrho}) \leq r$ (where again we indicate with $\epsilon_k(\cdot)$ the singular values of a matrix)~\cite{LiuRandomized}. Such a criterion appeared already in the literature~\cite{JohnstonKribs15}, and can be seen as a generalization of the renowned Computable Cross-Norm and Realignment (CCNR) criterion to detect the Schmidt number~\footnote{See also \cite{ChenPRL05,ZhangZhangZhangGuo2007} for other related criteria bounding the concurrence.}. Similarly, other criteria based on unitarily-invariant norms, can be shown to be weaker than simple corollaries of our \cref{theo1}. See also Ref.~\cite{LiuRandomized} for more details on such a comparison.

\subsection{Schmidt-correlated states}

One advantage of \cref{cor:1} is that it is a simple criterion which is invariant under a change of the local bases in the definition of $\Gamma_\varrho$. In general, in fact, it is not obvious which basis is more convenient to consider for the evaluation of \cref{theo1}. See \cref{app:LOBs} for a discussion. A special case in which this is actually clear is when 
the density matrix can be written as a mixture of states which share the same Schmidt bases $\varrho=\sum_{kl} \varrho_{kl} \kb{kk}{ll}$, so called Schmidt-correlated states~\cite{gittsovich08}. In that case, a canonical basis which is optimal is constructed precisely from an operator basis associated to the two Schmidt bases. 
For this particular class of states we can even observe that our theorem provides a necessary and sufficient condition. A sketch of the proof follows, while all technical details are given in \cref{app:SchCorrSt}.

\begin{observation}\label{obs:SchmidtCorrSuffNece}
A Schmidt-correlated state $\varrho=\sum_{ij} \varrho_{ij} \ket{ii}\bra{jj}$ satisfies $s(\varrho)> r$ {\it if and only if} it violates \cref{eq:CMCrankr}.
\end{observation}

{\it Proof.---}The ``if'' direction is clear from \cref{theo1}. The proof of the ``only if'' part consists of two steps. First, one observes that for Schmidt-correlated states the states $\ket{\psi^{(k)}_r}$ that provide the lower bound on the right-hand side of \cref{eq:CMCrankr} must have the same Schmidt bases as $\varrho$ (cf. \cref{lemma:PSapp} in \cref{app:SchCorrSt}). Second, one observes that whenever \cref{eq:CMCrankr} is satisfied, these $\ket{\psi^{(k)}_r}$ provide a decomposition of $\varrho$, which thus must satisfy $s(\varrho) \leq r$ by definition \eqref{eq:SNdefinition}. \qed

\subsection{States with symmetries}

Another important simplification of the evaluation of the CMC arises for states with symmetries. In particular, consider states that are invariant under a subgroup of unitaries of the form 
$U = U_a \otimes U_b$, namely density matrices $\varrho$ such that $U \varrho U^\dagger = \varrho$ for all $U\in G$ where $G$ is some subgroup of unitaries. 
Equivalently, one can introduce the projection operation (sometimes also called twirling) onto the group invariant states~\cite{VollbrechtWerner01}:  
\be\label{eq:twirling}
\mathcal P(\varrho):= \int_G {\mathrm d} U U \varrho U^\dagger = \varrho ,
\ee
where the integral is over the Haar measure of $G$ and the equality holds for the invariant states $\varrho$. 
If a state $\varrho^\prime$ is not invariant under $U\in G$, the output state $\mathcal P(\varrho^\prime)$ (which does not coincide with $\varrho^\prime$ in this case) will be invariant under $U\in G$.

Depending on the specific group $G$ (especially from its size), the $G$-invariant states might have a very simplified form, which also allows to evaluate entanglement measures
far more easily~\cite{VollbrechtWerner01}. 
One of the most typical examples is given by so-called {\it isotropic states}, which are invariant under all unitaries of the form $U\otimes U^*$.
As a result, the most general isotropic state has the form
\be\label{eq:isostatesgen}
\varrho_{\rm is} = \frac{1-\alpha}{d^2} \id \otimes \id + \alpha \kb{\psi_+}{\psi_+} = \frac{\id \otimes \id}{d^2} + \frac{\alpha}{d} \sum_{k=1}^{d^2-1} \sigma_k \otimes \sigma^*_k ,
\ee
where 
\be
\alpha  = \frac{d^2}{d^2-1}\tr\left( \kb{\psi_+}{\psi_+} \varrho_{\rm is} \right) - \frac{1}{d^2-1}  
\ee
is a real coefficient and $d$ is the dimension of the single particle space. Here $(\sigma_1, \dots , \sigma_{d^2-1})$ is any $su(d)$ basis.
As a consequence of such a big symmetry, as we mentioned, all calculations of entanglement measures on such states are greatly simplified. 
The same happens for the evaluation of the CMC.
Using the form in \cref{eq:isostatesgen}, one can readily calculate the elements of the covariance matrix $\Gamma_{\varrho_{\rm is}}$ in a standard pair of  
bases and one can get the condition in \cref{eq:fidcrit} as a corollary of the CMC in a very simple way. 
In particular, such highly symmetric states have a diagonal cross-covariance matrix as well as diagonal local covariance matrices (and moreover with all nonzero diagonal elements equal to each other in this case).
See \cref{app:isostates} for more details.

Note once more, that {\it all quantum states} can be projected/twirled into a state with any symmetry. In particular, for example every state can be projected into the set of isotropic states.
In this way, one can see in a simple way that the criterion in \cref{eq:fidcrit} follows as a corollary of the CMC: (i) Given any state $\varrho$, first one projects it into an isotropic state,
which is given by essentially only the parameter $\alpha$, i.e., essentially only by its fidelity with respect to the maximally entangled state; (ii) second, one calculates the covariance matrix of such a projected isotropic state and plugs the resulting traces of submatrices into \cref{eq:cor1}, which then becomes nothing but \cref{eq:fidcrit}.

\subsection{Spin states}

In general, the criterion \cref{eq:CMCrankr} is a compact way of enclosing many individual conditions 
in the form of variance-based uncertainty relations 
\be
\sum_k \var{T_k}_\varrho \geq \min_{\psi_r} \sum_k \var{T_k}_{\psi_r} ,
\ee
with $T_k=A_k\otimes \id + \id \otimes B_k$ being collective observables. 
In fact, a weaker version of \cref{eq:cor1} is 
\be 
\tr(\varrho^2_a)+\tr(\varrho^2_b)+2\tr|X_{\varrho}| \leq 2r ,
\ee 
which is nothing but an uncertainty relation criterion for a full basis set, i.e., with $T_k^\pm = g_k^A \otimes \id \pm \id \otimes g_k^B$, the bound being $2(d-r)$. Note also that a Schmidt-number criterion with a similar expression already appeared in the literature~\cite{deVicentePRAr07}, however with a strictly worse bound than our criterion.

Another paradigmatic case consists of spin measurements with spin quantum number $j_n=(d_n-1)/2$ along three orthogonal directions $\obsset{j}_n=(j^{(n)}_x, j^{(n)}_y, j^{(n)}_z)$, which is what is typically measured in, e.g., cold-atom experiments~\cite{GuheneToth09,FriisNatPhys19}. 
Considering now the covariance matrix of such spin components $\Cov_\varrho(\obsset j)$, and applying essentially the same reasoning as in the proof of \cref{cor:1}, one could be able to derive the following family of inequalities:
\be\label{eq:spinIneqPR}
\av{\vec j_a}^2_\varrho + 4t^2 \av{\vec j_b}^2_\varrho +4 |t| \tr|X^{(j)}_\varrho|
\leq \mathcal{B} (t),
\ee
where $\av{\obsset j_n}^2:=\av{j^{(n)}_x}^2+\av{j^{(n)}_y}^2+\av{j^{(n)}_z}^2$, and 
\be\label{eq:genboundB}
\mathcal{B} (t):= \max_{\psi_r} \left[ \av{\vec j_a}^2_{\psi_r} + 4t^2 \av{\vec j_b}^2_{\psi_r} +4 |t| \tr|X^{(j)}_{\psi_r}| \right] ,
\ee
with the maximum taken over all pure Schmidt-rank-$r$ states\footnote{Note that this expression leads in general to bounds which are tighter than simply taking a bound like $\tilde{\mathcal{B}} (t) = j_a^2 + 4t^2 j_b^2 +4|t| \max_{\psi_r} \tr|X^{(j)}_{\psi_r}|\geq \mathcal B(t)$. In fact, this last one is a trivial bound and does not lead to useful Schmidt-number criteria.}.

A similar reasoning can be applied to any set of observables, i.e., finding a bound on the the analogous expression can be used to derive an inequality similar to the above (and to \cref{eq:cor1}). Of course, the difficulty lies on finding such a bound, which in general is hard to do analytically, but it can be handled numerically, at least for small values of $r$.
Here we present a simplified expression for the bound, which is valid only for the subset of Schmidt-correlated states, see \cref{app:spin}. 
However, preliminary numerical investigation indicates that such a bound could be valid for all states, and also that the most useful inequality in \cref{eq:spinIneqPR} is obtained for $t=1/2$.
In this case the expression can be rewritten in the form 
\be\label{eq:spinineexpr}
\var{J_x^\pm}+\var{J_y^\pm}+\var{J_z^\pm} \geq \B_{r,j_a,j_b}(1/2) ,
\ee
where $J_k^\pm = j_k\otimes \id \pm \id \otimes j_k$ are collective spin observables and $\B_{r,j_a,j_b}(1/2)$ follows from the corresponding $\mathcal B(1/2)$.

In the case of equal dimensions such a bound reads 
\be\label{eq:Kboundonehalf}
K_{r,j}(1/2) = \min_{\vec \lambda} \sum_{k=1}^{2j} k(2j+1-k)\left(\sqrt{\lambda_k}-\sqrt{\lambda_{k+1}}\right)^2 ,
\ee
and the minimization is taken over all probability vectors $\vec \lambda$ with only $r$ nonzero components. This minimum is easy to find numerically, and for $r=2$ can be even calculated analytically for all $j$ and reads 
\be
\B_{2,j}=4j-1-\sqrt{1-4j+8j^2}. 
\ee
Moreover further (non-optimal) bounds can be derived analytically for all $r$. These read 
\be\label{eq:anaboundj}
\B_{r,j} \geq \left( r\sum_{k=1}^{r} k^{-1}(2j+1-k)^{-1} \right)^{-1} \geq \frac{2j}{r^2}, 
\ee
and are proven in \cref{lemma:2} in \cref{app:spin}.
One can observe that mixed states in the vicinity of typical spin states prepared in experiments, such as singlet states, Dicke states and spin squeezed states are detected even by this weaker version of the inequality.
However, a preliminary numerical investigation suggests that the states detected by such an inequality have to be very close to the ideal pure state, hence with
a very low amount of noise. At the same time,
note that the case of $r=1$ is equivalent to the criterion based on local uncertainty relations \cite{hofman03}. Thus our inequality
can still be seen as a generalization of the latter for detecting the entanglement dimensionality of Schmidt-correlated states, which is at least an interesting theoretical observation that could lead to a future more practically useful condition for detecting the Schmidt number of spin-squeezed states. 

In this respect, we also made a numerical investigation on randomly generated spin states (not Schmidt-correlated) in dimension $d=3$ (i.e., $j=1$) and observed that our condition in \cref{eq:Kboundonehalf} remains valid. In particular, we observed that the optimal pure Schmidt-rank-$r$ states that would give the bound in \cref{eq:genboundB} have Schmidt bases that correspond to the $j_z$ eigenbases, which indeed seems quite natural at a first sight. 
Furthermore, we note that the expression on the left-hand side of \cref{eq:spinIneqPR} has some form of rotational invariance. In particular the expression on the left-hand side of \cref{eq:spinineexpr} with all operators $J_k^{+}$ is invariant under global rotations. Thus, the same expression could be obtained by first projecting the state $\varrho$ into a rotationally-invariant state, such as those discussed in \cref{app:rotinvstates}, which have a very simple expression for the spin covariance matrix. In that case we conjecture, also based on some very preliminary numerical calculations on $3$-dimensional systems, that the optimal states $\ket{\psi_r}$ leading
to the bound in \cref{eq:genboundB} can be chosen to have Schmidt bases aligned with, say, the local $j_z$ eigenbases.
So far, we have not been able to prove such a conjecture. Nevertheless, we believe that further investigation along these lines could lead to a more practical implementations of our method in realistic highly-entangled spin states, such as those produced, e.g., in cold atoms.

\subsection{Outlook: PPT states and multipartite extension}

An intriguing question to gauge the power of our method is whether it could be used to detect forms of entanglement which are usually thought as somehow hard to
detect. Perhaps the most prominent example is that of so-called PPT-entangled states, i.e., states that are not detected by \cref{eq:PPTcrit}. 
On the other hand, it has been recently found that there are classes of PPT states that have very high Schmidt number~\cite{HuberLamiLancienMuellerHermes2018}.
This class is a generalization of known example of PPT states that have been already shown to be in some sense highly entangled~\cite{IshizakaBound2004,PianiClass2007,LamiHuber2016} and are also considered in other contexts~\cite{tothPRA09,ImaiBound2021,hiesmayr2021free,Marcin2022twoqutrit,BenknerCharacterizing2022}.

It is worth noting that these constructions are typically of multipartite states that are PPT-entangled across a given bipartition. 
In particular, as noted in~\cite{HuberLamiLancienMuellerHermes2018}, one can make a general construction of multipartite states that have a high Schmidt number with respect to a certain bipartition and are also PPT with respect to the same bipartition. 
One important example is the following $4$-partite state with dimensions $d_{a_1} = d_{b_1}=d_1$ and $d_{a_2} = d_{b_2}=d_2$:
\be
\varrho_{a_1 b_1 a_2 b_2} = X_{a_1 b_1} \otimes \left( \id - \kb{\psi_+}{\psi_+} \right)_{a_2 b_2} + Y_{a_1 b_1} \otimes \kb{\psi_+}{\psi_+}_{a_2 b_2} ,
\ee
where the operators $X_{a_1 b_1}$ and $Y_{a_1 b_1}$ have to satisfy certain properties in order for the state to be PPT and of Schmidt number $s\geq \lceil d_2/d_1 \rceil$ with respect to the partition $(a_1 a_2 | b_1 b_2)$. 
One important example state from the class above has $d_{a_1} = d_{b_1} = 2$ and $d_{a_2} = d_{b_2} = d/2$ and is discussed in \cref{app:PPTstatesSchmidt}.

Thus, the question arises whether we can detect those states with our criterion, perhaps even with the simple \cref{cor:1}. Note, in fact, that these PPT entangled states
are also highly symmetric, namely they are invariant under unitaries of the type $U\otimes V \otimes U^* \otimes V^*$. See \cref{app:4partiUUVV} and \cref{app:PPTstatesSchmidt}. 

However, one can see that actually \cref{cor:1} does not work to detect a nontrivial Schmidt number in those states, and therefore further investigation is needed in order to check whether the full CMC in \cref{theo1} will detect the state or not.

Before concluding, at this point it is also interesting to briefly discuss an outlook on the application of our methods to the case of general multipartite states. 
In such a case the entanglement dimensionality can be calculated across any bipartition and the notion of Schmidt number can be extended to that of Schmidt-number vector~\cite{Huber_2013}.
In this case, the full block-covariance matrix for each bipartition should be constructed from local orthogonal bases for the bipartition, for example by taking tensor products of the single particle operators in each party. 
For example, let us consider a $4$-particle state where the local dimensions are all equal to $d$. One can consider the bipartitions $1|234$, $12|34$, $123|4$ and so on. For the first bipartition the local operators are of the form $\obsset{g}_1=\{g_k^{(1)}\}_{k=1}^{d^2}$ and $\obsset{g}_{234} = \{g_k^{(2)}\otimes g_l^{(3)} \otimes g_m^{(4)}\}_{k,l,m=1}^{d^2}$, while for the second they are $\obsset{g}_{12}=\{g_k^{(1)}\otimes g_l^{(1)} \}_{k,l=1}^{d^2}$ and $\obsset{g}_{34}=\{g_k^{(3)} \otimes g_l^{(4)}\}_{k,l=1}^{d^2}$. Because of this, the block-covariance matrix has to be calculated independently for each different bipartition.
Another approach can be instead to consider only single particle operators and thus for an $n$-particle system the covariance matrix $\Cov(\obsset{g}_1,\obsset{g}_2,\dots,\obsset{g}_n)$ as it was already pointed out in \cite{GittsovichMulti} for the CMC entanglement criterion. By using such a multipartite local covariance matrix the advantage is that the different bipartitions can be simply analyzed by taking the corresponding minors of the full covariance matrix. 
However one is not considering a full basis for each bipartition, but just a smaller set of observables.

\section{Conclusions}

In conclusion, generalizing the work of Ref.~\cite{guhnecova} we have derived a covariance matrix criterion for the entanglement dimensionality. We have also discussed weaker corollaries of such a general criterion and their application to detect general states, which improves over other typical entanglement dimensionality witnesses. 
Because of that, the main corollary we presented can be readily applied to current experiments with highly flexible measurements, outperforming current methods used to detect the entanglement dimensionality, such as fidelity witnesses~\cite{FriisNatPhys19,Frerot2023,BavarescoNatPhys18,Malik_2016,Herrera-ValenciaSrivastavPivoluskaHuberFriisMcCutcheonMalik2020}.

We have also discussed how to find conditions involving virtually arbitrary subset of local operators, focusing on the paradigmatic example of spin components. 
The latter are routinely measured in atomic ensembles, and collective spin variances are almost the only observables that can be measured in those experiments.
Thus, our results in that respect can serve as a basis for the development of a practical experimentally implementable entanglement-dimensionality-detection method
that could find applications in concrete many-body experiments, prominently in atomic ensembles. However, in that case further investigation is needed to potentially
improve the practical implementation of our ideas. 
The current drawbacks of our spin-squeezing-like conditions are that either one has to find the bound \eqref{eq:genboundB} with a potentially complex numerical calculation, or one has to improve our current non-optimal bounds in \cref{eq:Kboundonehalf,eq:anaboundj}, which formally are only valid for Schmidt-correlated states and seem to be not very noise-tolerant. 
Similarly, we leave for future work a detailed extension of our method to the multipartite scenario, for example to detect the Schmidt number vector, as well as
further interesting theoretical questions that are raised by the present work, e.g., whether it is possible to detect high Schmidt number in some exemplary PPT states, such as those introduced in \cite{HuberLamiLancienMuellerHermes2018}.

{\bf Acknowledgments.---}This work is supported by the National Natural Science Foundation of China (Grants No. 11975026, 12125402) and  the Innovation Program for Quantum Science and Technology (No. 2021ZD0301500). SL acknowledges financial support from the China Scholarship Council (Grant No. 202006010302). MF was supported by the Swiss National Science Foundation, and by The Branco Weiss Fellowship -- Society in Science, administered by the ETH Z\"{u}rich. GV acknowledges financial support from the Austrian Science Fund (FWF) through the grants P 35810-N and P 36633-N (Stand-Alone) and ZK 3 (Zukunftskolleg). MH acknowledges funding from the European Research Council (Consolidator grant 'Cocoquest' 101043705).

\appendix

\section{Appendix}\label{supplement}

In this appendix we provide some formal proofs and details corroborating the statements in the main text.

\subsection{The entanglement parameter can not be connected with the Schmidt rank}\label{app:tparbound}
The entanglement parameter is defined as \cite{GittsovichPRA10}
\begin{equation}
\mathcal{E}(\varrho)=1-V(\varrho)
\end{equation}
where the function $V(\varrho)$ is
\begin{equation}
V(\varrho)=\max _{t, \kappa_A, \kappa_B}\left\{t \leq 1: \gamma(\varrho)-t \kappa_A \oplus \kappa_B \geq 0\right\}
\end{equation}
and $\kappa_{A,B}$ are convex combinations of pure states. For pure states in the Schmidt decomposition, the maximum value of nonnegative $V$ can be computed by
\begin{equation}
V_{max}=\max_{\mathcal{P}} \min _{i<j} \frac{\left(\sqrt{\lambda_i}-\sqrt{\lambda_j}\right)^2}{p_i+p_j}, 1 \leq i<j \leq d
\end{equation}
where $\mathcal{P}$ goes over all possible probability distributions $\{p_k\}$. However, for entangled pure states of $d\geq 4$, the Schmidt rank cannot be detected by $V$. This can be simply seen via the following example. Consider the state
$$\frac{1}{\sqrt{2}}(\ket{00}+\ket{11}),$$ which has Schmidt rank $2$.
In this case, the coefficients are
$$
\begin{aligned}
\lambda_1&=\lambda_2=\frac{1}{\sqrt{2}} \\
\lambda_3&=\cdots=\lambda_d=0 .
\end{aligned}
$$
Let us now assume $V>0$. Then, we should have $$\frac{\left(\sqrt{\lambda_i}-\sqrt{\lambda_j}\right)^2}{p_i+p_j} \neq 0 $$ 
for a probability distribution $\{p_k\}$ and for all $1\leq i<j\leq d$. This implies that
$$
p_1+p_2=0
$$
and
$$p_i+p_j=0$$ 
should hold for any $3\leq i<j\leq d$. This is however in contradiction with $\sum_i p_i=1$. Therefore $V=0$ must be the lower bound for all Schmidt-rank-$2$ entangled states and therefore also for higher Schmidt rank (pure) states. 

\subsection{Local orthogonal bases from useful state decompositions}\label{app:LOBs}

Here we discuss what is the optimal choice of local orthogonal bases for the evaluation of the covariance matrix criterion.
For simplicity, we consider the case of two equal dimensions. The more general case can be straightforwardly obtained by essentially substituting $d$ with the smallest between the two dimensions and adding zeroes where needed for the larger dimension.
First, let us consider a generic pure state, which can be written as
$\ket{\psi}=\sum_k \sqrt{\lambda_k} \ket{k k}$ in its Schmidt decomposition.
This corresponds to a local unitary transformation of the state.
Considering the two Schmidt bases 
as the canonical bases for the Hilbert spaces 
we can write canonical orthonormal bases for the local observables as 
$g^{(n)}_{kl}=\kb{k}{l}$ for $1\leq k ,l\leq d$. Hermitian orthonormal bases can be constructed from these as follows:
\begin{equation}\label{eq:StandardBases}
\obsset{g}_n=\left(\{|k\rangle\langle k|\}_{k=1}^{d},\left\{\frac{1}{\sqrt{2}}(|k\rangle\langle l|+| l\rangle\langle k|) \right\}_{k=1,k<l}^{d},\left\{\frac{i}{\sqrt{2}}(|k\rangle\langle l|-| l\rangle\langle k|) \right\}_{k=1,k<l}^{d}\right) .
\end{equation}
For compactness, we can use a multi-index notation $\nu=(kl)$, with $\nu$ a $2$-digit $d$-nary number.
For instance, in the case $d=2$ this canonical basis is given by 
\begin{equation}
\left\{
\left(\begin{array}{cc}
1 & 0 \\
0 & 0
\end{array}\right),
\left(\begin{array}{cc}
0 & 0 \\
0 & 1
\end{array}\right),
\left(\begin{array}{cc}
0 & \frac{1}{\sqrt{2}} \\
\frac{1}{\sqrt{2}} & 0
\end{array}\right),
\left(\begin{array}{cc}
0 & \frac{i}{\sqrt{2}} \\
-\frac{i}{\sqrt{2}} & 0
\end{array}\right)
\right\}.
\end{equation}

In such canonical bases, the covariance matrix of the generic $\ket{\psi_r}$ contains blocks of the form
$\gamma_a=\gamma_b=$ $\operatorname{diag}(D, R, I)$ and $X_r=\operatorname{diag}\left(D, R_C, I_C\right)$ with $D_{k l}=\lambda_k \delta_{kl}-\lambda_k \lambda_l$ being a $(d \times d)$ matrix and $R=I=\operatorname{diag}\left\{\tfrac 1 2 (\lambda_k+\lambda_l)\right\}_{k<l=1}^{d}$ being diagonal matrices of dimension $d(d-1) / 2$ and similarly $R_C=-I_C=\operatorname{diag}\left\{\sqrt{\lambda_k \lambda_l}\right\}_{k<l=1}^d$. 
Thus, we see that, apart from the block $D$ corresponding to the diagonal operators in all blocks, the covariance matrix is block-diagonal and we can read off the singular values of the blocks.
Note, in fact, that a change of local bases $\obsset{g}$ corresponds to a local orthogonal transformation $O_a \oplus O_b$ at the level of the covariance matrix and thus the most general covariance matrix for the state is obtained by applying orthogonal transformations on its blocks as
\be\label{eq:genCovPure}
\Cov_\psi(\obsset{g}) = 
\left(\begin{array}{ll}
O_a \gamma_a O_a^T & O_a X_\psi O_b^T \\
O_b X_\psi^T O_a^T & O_b \gamma_b O_b^T
\end{array}\right) .
\ee
In particular, we get that for a Schmidt-rank-$r$ pure state the nonzero singular values of the covariance matrix blocks are as in \cref{eq:kappaCconds} in the main text.

Let us now consider a mixed state $\varrho$. With a local orthogonal change of bases we can now bring the density matrix to the Schmidt decomposition in operator basis:
\be
\varrho = \sum_k \xi_k g^{(a)}_k \otimes g^{(b)}_k ,
\ee
with $\xi_k\geq 0$, which is obtained from the singular value decomposition of the density matrix expansion in a generic pair of bases $\varrho = \sum_{kl} C_{kl} (g^\prime)^{(a)}_k \otimes (g^\prime)^{(b)}_l$. 
In this bases the cross-correlation matrix is in its singular value decomposition, namely $\av{g^{(a)}_k \otimes g^{(b)}_l}_\varrho = \xi_k \delta_{kl}$ and the $\xi_k$ are positive. 
However this is not the case for the covariances. In particular, in these two Schmidt operator-bases the covariance matrix has blocks given by
\be
\begin{aligned}
(\gamma_a)_{kl} &=  \sum_n \xi_n \tr\left(\frac{g_k^{(a)} g^{(a)}_l+g_l^{(a)} g^{(a)}_k}{2} g_n^{(a)}\right) \tr(g^{(b)}_n) - \xi_k \xi_l \tr(g^{(b)}_k)\tr(g^{(b)}_l) , \\
(\gamma_b)_{kl} &=  \sum_n \xi_n \tr(g^{(a)}_n) \tr\left(\frac{g_k^{(b)} g^{(b)}_l+g_l^{(b)} g^{(b)}_k}{2} g_n^{(b)}\right) - \xi_k \xi_l \tr(g^{(a)}_k)\tr(g^{(a)}_l)  , \\
(\C)_{kl} &= \xi_k \delta_{kl} - \xi_k \xi_l \tr(g^{(b)}_k) \tr(g^{(a)}_l)  ,
\end{aligned}
\ee
where we used the orthonormality relations $\tr(g_k g_l)=\delta_{kl}$ which hold for both bases.

Another choice could be to consider the bases in which the cross-covariances are in the singular value form, namely the bases $\obsset{\tilde g}$ such that $\av{\tilde g^{(a)}_k \otimes \tilde g^{(b)}_l}_\varrho - \av{\tilde g^{(a)}_k}_\varrho \av{\tilde g^{(b)}_l}_\varrho = \zeta_k \delta_{kl}$ and the $\zeta_k$ are positive. Note that since the covariances of $\id\otimes g$ and $g\otimes \id$
are always zero for any general observable $g$ we have that the bases $\obsset{\tilde g}_n$ for the two parties are of the form $\obsset{\tilde g}_n = (\id /\sqrt{d_n}, \obsset{\sigma}_n)$ where $\obsset{\sigma}_n$ are $su(d_n)$ bases for the two parties $n=a,b$. 
Bases for which both the cross-correlations and the cross-covariances are diagonal are obtained from the so-called {\it Filtered Normal Form} (FNF) of a density matrix, which is:
\be
\varrho_{\rm FNF} = \tfrac 1 {d_a d_b}\id \otimes \id + \sum_{k=1}^{d^2-1} \xi_k \sigma_k^{(a)} \otimes  \sigma_k^{(b)} ,
\ee
where again $\obsset{\sigma}_a=(\sigma_1^{(a)},\dots,\sigma_{d_a^2-1}^{(a)})$ and $\obsset{\sigma}_b=(\sigma_1^{(b)},\dots,\sigma_{d_b^2-1}^{(b)})$ are bases of the special unitary algebra and in particular are traceless (in the formula above we called $d$ the smallest between the two dimensions).
Such a FNF exists for every density matrix of full rank and moreover it is not unique. 
To bring a state in such a form, one needs to perform local linear transformations which are not unitary, i.e.,  
$\varrho_{\rm FNF} = (F_a \otimes F_b) \varrho (F_a \otimes F_b)^\dagger$ with $F_a \in SL(d_a,\mathbb C)$ and $F_b \in SL(d_b,\mathbb C)$ which are called local filtering operations.
These operations cannot map separable states into entangled ones nor vice-versa, but can change the value of entanglement monotones. In fact, it is known that in some cases bringing a state in such a form maximizes the value of entanglement monotones~\cite{VerstraeteDehaeneDeMoor2003}.
However, it has been shown~\cite{gittsovich08} that, although in most cases bringing a state in the FNF helps to detect it with the CMC, there are some states for which this is not the case, namely the FNF is not detected while the original state is.

\subsection{Schmidt-correlated states}\label{app:SchCorrSt}

Here we prove that for so-called Schmidt-correlated states, i.e., density matrices that have a decomposition in pure states all sharing the same Schmidt bases, the CMC criterion is both necessary and sufficient to detect the Schmidt number. First, we need the following lemma.

\begin{lemma}\label{lemma:PSapp}
Let us consider a Schmidt-correlated density matrix $\varrho=\sum_{ij} \varrho_{ij} \kb{ii}{jj}$ and the associated covariance matrix $\Gamma_{\varrho}$ in the canonical operator basis constructed from its Schmidt bases.
The matrix $\DiffMat:= \Gamma_{\varrho} - \sum_k p_k \Gamma_{\psi_k}$ as in \cref{eq:CMCrankr} in the main text can be positive semidefinite only if all states $\ket{\psi_k}$ have the same Schmidt bases as $\varrho$.
\end{lemma}

{\it Proof.---}Let us consider generic states $\ket{\psi_{k}}=\sum_{i,j} m_{ij}^{(k)}|ij\rangle$, expanded in the Schmidt bases of $\varrho$ and let us consider the principal minor of $\DiffMat$ corresponding to the operators $\{\ket{i}\bra{i}\otimes\mathbb{1},\mathbb{1}\otimes\ket{i}\bra{i}\}$:
\begin{equation}
\DiffMat_{(\nu \nu)}=\left(\begin{array}{ll}
D_{ii}-\sum_{k}p_{k}a_{k,i} & D_{ii}-\sum_{k}p_{k}c_{k,i} \\
D_{ii}-\sum_{k}p_{k}c_{k,i} & D_{ii}-\sum_{k}p_{k}b_{k,i}
\end{array}\right),\text{ with } \nu=(ii), D_{i i}=\varrho_{ii} - \varrho_{ii}^2,
\end{equation}
where the generic elements $a_{k,i}, b_{k,i}, c_{k,i}$ are given by: $a_{k,i}=\sum_{j}|m_{ij}^{(k)}|^2-\left(\sum_{j}|m_{ij}^{(k)}|^2\right)^2$, $b_{k,i}=\sum_{j}|m_{ji}^{(k)}|^2-\left(\sum_{j}|m_{ji}^{(k)}|^2\right)^2$ and $c_{k,i}=|m_{ii}^{(k)}|^2-\left(\sum_{j}|m_{ij}^{(k)}|^2\right) \left(\sum_{j}|m_{ji}^{(k)}|^2\right)$. 
From the positivity of $\Gamma_{\psi_k}$ we have  \begin{equation}
\label{eq:lemma3Coef}
\sqrt{\left(\sum_{k} p_{k} a_{k,i}\right)\left(\sum_{k} p_{k} b_{k,i}\right)} \geqslant \sum_{k} p_{k} \sqrt{a_{k,i} b_{k,i}} \geqslant \sum_{k} p_{k} c_{k,i}
\end{equation}
Then notice that $\DiffMat_{(\nu \nu)}$ is positive only when $\sum_{k} p_{k} a_{k,i}=\sum_{k} p_{k} b_{k,i}=\sum_{k} p_{k} c_{k,i} \leqslant D_{i i}$. In this case \cref{eq:lemma3Coef} saturates, which means that $a_{k,i}=b_{k,i}=c_{k,i}$ for all $k$. Imposing that $a_{k,i}=b_{k,i}$ and $a_{k,i}=c_{k,i}$ gives
$m^{(k)}_{i j}=0$ for all $j \neq i$. This means that $\ket{\psi_{k}}=\sum_{i} m^{(k)}_{ii} |ii\rangle$ for all $k$, which proves the claim.
\qed

With this, we are now in the position to prove \cref{obs:SchmidtCorrSuffNece}.

{\it Proof of \cref{obs:SchmidtCorrSuffNece}.---}Let us consider a density matrix of the form $\varrho=\sum_{ij} \varrho_{ij} \ket{ii}\bra{jj}$. The coefficients $\varrho_{ij}$ can be written as $\varrho_{i j}=\sum_{u} q_{u} \sqrt{\Lambda_{i}^{(u)} \Lambda_{j}^{(u)}}$ where $\Lambda_{i}^{(u)}$ are Schmidt coefficients of every pure state component of $\varrho$. The blocks in the covariance matrix of $\varrho$ are given by $\gamma_{a}=\gamma_{b}=\operatorname{diag}(D, R, I)$ and $X_{\varrho}=\operatorname{diag}\left(D, R_{C}, I_{C}\right)$ with $D_{jk}=\varrho_{jj} \delta_{jk}-\varrho_{jj} \varrho_{kk}$, $R=I=\frac{1}{2} \operatorname{diag}\left\{\varrho_{jj}+\varrho_{kk}\right\}_{j<k=1}^{d}$ and $R_{C}=-I_{C}=\operatorname{diag}\left\{\varrho_{jk}\right\}_{j<k=1}^{d}$. Using \cref{lemma:PSapp} above, we know that \cref{eq:CMCrankr} can be satisfied only if all the $\ket{\psi^{(k)}_r}$ share the same Schmidt bases as $\varrho$.  Let us consider the principal minor of $\DiffMat=\Gamma_{\varrho}-\sum_{k} p_{k} \Gamma_{r}^{(k)}$ corresponding to the operators $\left\{\frac{1}{\sqrt{2}}(|j\rangle\langle k|+| k\rangle\langle j|) \otimes \mathbb{1}, \mathbb{1}\otimes\frac{1}{\sqrt{2}}(|j\rangle\langle k|+| k\rangle\langle j|)\right\}$:
\begin{equation}
\DiffMat_{(\nu \nu)}=\left(\begin{array}{ll}
R_{\nu \nu}-\sum_{u} p_{u} R_{\nu \nu}^{(u)} & (R_{C})_{\nu \nu}-\sum_{u} p_{u} (R_{C}^{(u)})_{\nu \nu} \\
(R_{C})_{\nu \nu}-\sum_{u} p_{u} (R_{C}^{(u)})_{\nu \nu} & R_{\nu \nu}-\sum_{u} p_{u} R_{\nu \nu}^{(u)}
\end{array}\right),\text{ with }\nu=(jk),j<k.
\end{equation}
From the positivity of $\DiffMat_{(\nu \nu)}$ we have $R_{\nu \nu}-\sum_{u} p_{u} R_{\nu \nu}^{(u)}\geq 0$ for all $\nu$. Since $\sum_{j<k}R_{jk,jk}=\frac{1}{2}(d-1)\sum_{j}\varrho_{jj}=\frac{1}{2}(d-1)$ and $\sum_{j<k}R_{jk,jk}^{(u)}=\frac{1}{2}(d-1)\sum_{j}\lambda_{j}=\frac{1}{2}(d-1)$, then we have that $\DiffMat_{(\nu \nu)}$ is the null matrix and therefore $\varrho_{jj}=\sum_{u} p_{u} \lambda_{j}^{(u)}$ and $\varrho_{jk}=\sum_{u} p_{u} \sqrt{\lambda_{j}^{(u)} \lambda_{k}^{(u)}}$. By setting $\Lambda=\lambda$ and $q=p$, we can then find a set of pure states $\ket{\psi_{r}^{(k)}}$ which satisfy \cref{eq:CMCrankr}, which also provides a decomposition of $\varrho$. In summary, if \cref{eq:CMCrankr} is violated, it means one cannot find a decomposition using $\ket{\psi_{r}^{(k)}}$ whose Schmidt rank is smaller than or equal to $r$, i.e., the Schmidt number of $\varrho$ is greater than $r$. Conversely, if the Schmidt number of $\varrho$ is greater than $r$, then one cannot have $\varrho_{jj}=\sum_{u} p_{u} \lambda_{j}^{(u)}$ and $\varrho_{jk}=\sum_{u} p_{u} \sqrt{\lambda_{j}^{(u)} \lambda_{k}^{(u)}}$, which means that \cref{eq:CMCrankr} cannot be satisfied. 
In fact, whenever these last equations are satisfied, then one can find a decomposition of $\varrho$ in terms of Schmidt-rank-$r$ pure states.
In conclusion, the violation of \cref{eq:CMCrankr} is necessary and sufficient for $s(\varrho) > r$ to hold.
\qed

\subsection{Inequality with the spin operators}\label{app:spin}

Here we consider the covariance matrix of three spin operators $\obsset j_n = (j^{(n)}_x,j^{(n)}_y,j^{(n)}_z)$ for the two parties $n=a,b$. We construct it by expressing the local spin operators in our chosen canonical bases, i.e., $j^{(n)}_x = \sum_k x^{(n)}_k g^{(n)}_k$, $j^{(n)}_y = \sum_k y^{(n)}_k g^{(n)}_k$ and $j^{(n)}_z = \sum_k z^{(n)}_k g^{(n)}_k$ and then construct the spin covariances by multiplying our covariance matrix with the vectors $\vec x_n = (x^{(n)}_1,\dots ,x^{(n)}_{d_n^2})$, $\vec y_n = (y^{(n)}_1,\dots ,y^{(n)}_{d_n^2})$ and $\vec z_n = (z^{(n)}_1,\dots ,z^{(n)}_{d_n^2})$ as
\be\label{eq:SpinCovMatGen}
\begin{aligned}
\Cov_{\varrho}(\obsset{j}_n)&= \left(\begin{array}{ccc}
\vec x_n \gamma_n \vec x_n^T & \vec x_n \gamma_n \vec y_n^T & \vec x_n \gamma_n \vec z_n^T \\
\vec y_n \gamma_n \vec x_n^T & \vec y_n \gamma_n \vec y_n^T & \vec y_n \gamma_n \vec z_n^T \\
\vec z_n \gamma_n \vec x_n^T & \vec z_n \gamma_n \vec y_n^T & \vec z_n \gamma_n \vec z_n^T 
\end{array}\right) \\
X^{(j)}&= \left(\begin{array}{ccc}
\vec x_a X_\varrho \vec x_b^T & \vec x_a X_\varrho \vec y_b^T & \vec x_a X_\varrho \vec z_b^T \\
\vec y_a X_\varrho \vec x_b^T & \vec y_a X_\varrho \vec y_b^T & \vec y_a X_\varrho \vec z_b^T \\
\vec z_a X_\varrho \vec x_b^T & \vec z_a X_\varrho \vec y_b^T & \vec z_a X_\varrho \vec z_b^T 
\end{array}\right) .
\end{aligned}
\ee
This way we can apply essentially the same reasoning as in \cref{theo1} and \cref{cor:1} to find Schmidt-number criteria involving only local spin measurements. 

In particular, we use the fact that the matrix $\DiffMat := \Cov_{\varrho}(\obsset{j}) - \sum_k p_k \Cov_{\psi_r}^{(k)}(\obsset{j})$ is also a positive matrix for all Schmidt-number-$r$ states. Again, this implies a condition analogous to \cref{eq:tangents} for any unitarily-invariant norm, and in particular the trace norm $\| A \| = \tr|A|$. Thus, we obtain
\be
\tr(\DiffMat_a) + 4 t^2 \tr(\DiffMat_b) \geq 4 |t| (\tr|X^{(j)}_\varrho| - \tr|X^{(j)}_{\psi_r}|) , 
\ee
where we indicated as $X^{(j)}$ the cross-covariance matrix of the spin operators.
Taking an optimal pure state $\ket{\psi_r}$ we can also drop the probabilities due to $\sum_k p_k=1$.
Considering first the single-party principal minors we get
\be
\tr(\DiffMat_a)= \sum_{k=x,y,z} \var{j^{(a)}_k}_\varrho - \var{j^{(a)}_k}_{\psi_r} = \av{\vec j_a}^2_{\psi_r} - \av{\vec j_a}^2_\varrho , 
\ee
where we 
called $\av{\vec j_a}^2 = \sum_{k=x,y,z} \av{j^{(a)}_k}^2$ and
used that $\sum_{k=x,y,z} (j_k^{(a)})^2 = j_a(j_a+1) \id$ 
with $j_a=(d_a-1)/2$ being the spin quantum number of party $a$. The same reasoning can be made for party $b$.
This way we get to the following inequality
\be\label{eq:partialBoundSpin}
\av{\vec j_a}^2_\varrho + 4t^2 \av{\vec j_b}^2_\varrho +4 |t| \tr|X^{(j)}_\varrho|
\leq \av{\vec j_a}^2_{\psi_r} + 4t^2 \av{\vec j_b}^2_{\psi_r} +4 |t| \tr|X^{(j)}_{\psi_r}| . 
\ee
Thus, now what is missing would be a further bound on the right-hand side, namely one should perform the following maximization
\be\label{eq:maxPsirSpinGen}
\max_{\psi_r} \left[ \av{\vec j_a}^2_{\psi_r} + 4t^2 \av{\vec j_b}^2_{\psi_r} +4 |t| \tr|X^{(j)}_{\psi_r}| \right] := \mathcal B(t) ,
\ee
which is performed over all pure states of fixed Schmidt rank equal to $r$ and
at fixed $t\geq 0$.

To make such a maximization in practice, one can start from a generic expression for the covariance matrix of a pure state, as in \cref{eq:genCovPure}, restricted to the case in which $O_a$ and $O_b$ come from a unitary transformation on the state. Afterwards, one can construct the spin covariance matrix as in \cref{eq:SpinCovMatGen} and calculate the expression in \cref{eq:maxPsirSpinGen}. This will be parametrized by the coefficients of the generic local unitary transformations plus the (generic) Schmidt coefficients of the state.

However there is a special case in which we can actually compute even an analytic upper bound.
This is in the following situation. Let us take our computational bases to be the eigenbases of $j_z^{(a)}$ and $j_z^{(b)}$. If our density matrix is Schmidt-correlated in our computational bases, namely we have $\varrho = \sum_{jk} \varrho_{jk} \kb{jj}{kk}$, then a straightforward calculation shows that the spin cross-covariance matrix is diagonal.
In fact, we can observe this by taking the full covariance matrix in the canonical bases \eqref{eq:StandardBases} and express the spin operators as
\begin{equation}
\begin{aligned}
j_x & =\sum_{l<k} \frac{1}{\sqrt{2}} \sqrt{\left(j(j+1)-m_l\left(m_l+1\right)\right)} \delta_{l+1, k} \frac{1}{\sqrt{2}}(|k\rangle\langle l|+| l\rangle\langle k|) \quad \Rightarrow \quad 
x_{l<k}^{(n)}=w_l \  \delta_{l+1, k}\\
\\
j_y & =\sum_{l>k} \frac{1}{\sqrt{2}} \sqrt{\left(j(j+1)-m_k\left(m_k+1\right)\right)} \delta_{l, k+1} \frac{i}{\sqrt{2}}(|k\rangle\langle l|-| l\rangle\langle k|) \quad \Rightarrow \quad 
y_{l>k}^{(n)}=w_k \  \delta_{l, k+1}\\
 \\
j_z & =\sum_k m_k \ket{k}\bra{k} \quad \Rightarrow \quad 
z_{l=k}^{(n)}=m_k ,
\end{aligned}
\end{equation}
with $-j\leq m_k \leq j$ being the $j_z$ eigenvalues and $w_k=\tfrac{1}{\sqrt{2}} \sqrt{\left(j(j+1)-m_k\left(m_k+1\right)\right)}$.
For simplicity, in the following we choose the ordering of the eigenvalues of both $j_z^{(a)}$ and $j_z^{(b)}$ to be increasing in our computational basis. 
In other words, we choose $m_k =k-1-j$ for both parties $a$ and $b$.

In this case, calling $\delta_a=\frac{1}{2} \sum_{i=1}^{d_a-1}\left(\varrho_{i i}+\varrho_{i+1, i+1}\right) (w_i^a)^2$, $\delta_b =\frac{1}{2} \sum_{i=1}^{d_a-1}\left(\varrho_{i i}+\varrho_{i+1, i+1}\right)(w_i^b)^2+\frac{1}{2} \varrho_{d_a d_a}(w_i^b)^2$ and $\delta_{ab}=\sum_{i=1}^{d_a-1} \varrho_{i, i+1} w_i^a w_i^b$ we get
\be
\begin{aligned}\label{eq:SpinCovsSCstates}
\Cov_{\varrho}(\obsset{j}_a)&=\diag\left\{\delta_a,\delta_a,
\sum_{i=1}^{d_a} \varrho_{i i} (m_i^{a})^2-\left(\sum_{i=1}^{d_a} \varrho_{i i} m_i^a\right)^2\right\} \ , \\
\Cov_{\varrho}(\obsset{j}_b)&=\diag\left\{\delta_b,\delta_b,
\sum_{i=1}^{d_b} \varrho_{i i} (m_i^{b})^2-\left(\sum_{i=1}^{d_a} \varrho_{i i} m_j^b\right)^2\right\} \ , \\
X^{(j)}&=\diag\left\{\delta_{ab},-\delta_{ab},\sum_{i=1}^{d_a} \varrho_{i i} m_i^a m_i^b-\left(\sum_{i=1}^{d_a} \varrho_{i i} m_i^a\right)\left(\sum_{i=1}^{d_a} \varrho_{i i} m_i^b\right)\right\} \ , 
\end{aligned}
\ee
assuming $j_a\leq j_b$ without loss of generality.

Furthermore, from \cref{lemma:PSapp} we know that in this case the optimal pure states $\ket{\psi_r}$ that provide the lower bound for the covariance matrix must have the same Schmidt bases, i.e., their Schmidt bases must coincide with our $j_z^{(n)}$ eigenbases. Thus, in this case we can write $\ket{\psi_r}=\sum_k \sqrt{\lambda_k}\ket{m_k^a m^b_k}$ 
and the spin covariance matrices for the optimal $\ket{\psi_r}$ will have the form as in \cref{eq:SpinCovsSCstates} 
with $\varrho_{jk} = \sqrt{\lambda_j \lambda_k}$, and $\lambda_k$ being their Schmidt coefficients.
In this particular case we can simplify the optimization in \cref{eq:maxPsirSpinGen} and obtain
\be\label{eq:maxPsirSpinSCstates}
\max_{\psi_r} \left[ \av{\vec j_a}^2_{\psi_r} + 4t^2 \av{\vec j_b}^2_{\psi_r} +4 |t| \tr|X^{(j)}_{\psi_r}| \right] \leq j_a\left(j_a+1\right)+4|t|^2 j_b\left(j_b+1\right) -  \B_{r,j_a,j_b}(t) , 
\ee
where we defined 
\be
\begin{aligned}\label{eq:BoundBGen}
\B_{r,j_a,j_b}(t) = \min_{\vec \lambda} \left\{ \sum_k \left[ \left(\sqrt{\lambda_k} w_k^a-2|t| \sqrt{\lambda_{k+1}} w_k^b \right)^2 + \left(\sqrt{\lambda_{k+1}} w_k^a -2|t| \sqrt{\lambda_k} w_k^b\right)^2 \right] \right\} ,
\end{aligned}
\ee
and $\vec \lambda$ is a vector with only $r$ nonzero entries which are positive and sum up to one. 
Again, this bound should be calculated at each different (real and positive) $t$.
The condition in \cref{eq:maxPsirSpinSCstates} is a consequence of the fact that 
\be
\begin{aligned}
&\left[ \av{\vec j_a}^2_{\psi_r} + 4t^2 \av{\vec j_b}^2_{\psi_r} +4 |t| \tr|X^{(j)}_{\psi_r}| \right] \\
\leq & \sum_{k} \left[ \lambda_k \left( (m_k^a)^2 +4t^2 (m_k^b)^2 \right) + 8 |t| \sqrt{\lambda_k \lambda_{k+1}} w_k^a w_k^b \right] \\
= & j_a\left(j_a+1\right)+4|t|^2 j_b\left(j_b+1\right)- \sum_k\left(\sqrt{\lambda_k} w_k^a-2|t| \sqrt{\lambda_{k+1}} w_k^b \right)^2 - \sum_k \left(\sqrt{\lambda_{k+1}} w_k^a -2|t| \sqrt{\lambda_k} w_k^b\right)^2
\end{aligned}
\ee
holds.
This in turn is due to the fact that either
\be
\av{\vec j_a}^2_{\psi_r} + 4t^2 \av{\vec j_b}^2_{\psi_r} +4 |t| |(X_{\psi_r}^{(j)})_{33}| 
= \left(\sum_m \lambda_m (m_a - 2|t| m_b)\right)^2 
+ 4|t| \left( \sum_m \lambda_m m_a m_b \right) 
\ee
or 
\be
\av{\vec j_a}^2_{\psi_r} + 4t^2 \av{\vec j_b}^2_{\psi_r} +4 |t| |(X_{\psi_r}^{(j)})_{33}| 
= \left(\sum_m \lambda_m (m_a + 2|t| m_b)\right)^2 
- 4|t| \left( \sum_m \lambda_m m_a m_b \right) 
\ee
and in both cases the expression is upper bounded by $\sum_m \lambda_m (m_a^2 + m_b^2)$ due to the convexity of the square, i.e.: $\left(\sum_m \lambda_m (m_a \pm 2|t| m_b)\right)^2 \leq \sum_m \lambda_m (m_a \pm 2|t| m_b)^2$ which is valid because $\{\lambda_m\}$ is a probability distribution.

The optimization in \cref{eq:BoundBGen} is relatively easy to perform numerically for every $t$. In the particular case
$t=1/2$ we can also derive an analytic bound to it as follows.

\begin{lemma}\label{lemma:2}
When $j_a=j_b=j,t=\frac{1}{2}$ and $r<d$, the bound $\B_{r,j}(t=1/2)$ satisfies
\be
\begin{aligned}
\B_{r,j}(1/2)&= \min_{\vec \lambda} \sum_k 2 w_k^2\left(\sqrt{\lambda_k}-\sqrt{\lambda_{k+1}}\right)^2 \geq \left( r\sum_{k=1}^{r} k^{-1}(2j+1-k)^{-1} \right)^{-1} \geq \frac{2j}{r^2} .
\end{aligned}
\ee
Furthermore, for $r=2$ we have the exact value
\be
\B_{2,j}(1/2)=4j-1-\sqrt{1-4j+8j^2} .
\ee
\end{lemma}

{\it Proof.---}By using Cauchy–Schwarz inequality, we have
\be
\begin{aligned}
\min_{\vec \lambda} \sum_{k=1}^{r} 2 w^2_k \left(\sqrt{\lambda_{k}}-\sqrt{\lambda_{k+1}}\right)^{2} &\geq \left(\sum_{k=1}^{r}(2 w^2_k)^{-1}\right)^{-1} \min_{\vec \lambda} \left(\sum_{k=1}^{r} \left|\sqrt{\lambda_{k}}-\sqrt{\lambda_{k+1}}\right|\right)^{2} \\
&\geq \frac{1}{r} \left(\sum_{k=1}^{r}(2 w^2_k)^{-1}\right)^{-1} , \quad \text{with} \quad 2 w^2_k=k(2j+1-k).
\end{aligned}
\ee
The last bound is because $\sum_{k=1}^{r}\left|\sqrt{\lambda_{k}}-\sqrt{\lambda_{k+1}}\right|$ reaches the minimum when $\lambda_{1}\geq\lambda_{2}\geq\cdots\geq\lambda_{r}$ and $\lambda_{1}\geq\frac{1}{r}$. Then we have 
\be
\begin{aligned}
\min_{\vec \lambda} \sum_{k=1}^{r} 2 w^2_k \left(\sqrt{\lambda_{k}}-\sqrt{\lambda_{k+1}}\right)^{2}
&\geq \frac{1}{r}\left(\sum_{k=1}^{r}\left(k(2j+1-k)\right)^{-1}\right)^{-1} \\
&\geq \left(r\sum_{k=1}^{r}(2j)^{-1}\right)^{-1} \\
&=\frac{2 j}{r^{2}} .
\end{aligned}
\ee
When $r=2<d$, the lowest value of $\sum_k 2 w_k^2\left(\sqrt{\lambda_k}-\sqrt{\lambda_{k+1}}\right)^2$ is 
\be
\min_{\lambda_{1}} 2j\left(\sqrt{\lambda_1}-\sqrt{1-\lambda_1}\right)^2+(4j-2)(1-\lambda_1)=4j-1-\sqrt{1-4j+8j^2} .
\ee
\qed

\subsection{States with symmetries}\label{app:symmstates}

Here, following the discussion in the main text, we consider states that are invariant under the twirling operation in \cref{eq:twirling}
for some example subgroup of local unitaries $U=U_a \otimes U_b$.

\subsubsection{Isotropic states}\label{app:isostates}
First, let us consider the isotropic states in \cref{eq:isostatesgen}, which are symmetric under unitaries of the form $U\otimes U^*$. Given their symmetry, we consider two basis of the standard form
\be
\obsset{g}_a=\left(\frac{\id}{\sqrt{d}} , \obsset{\sigma} \right) = \obsset{g}^*_b ,
\ee
where $\obsset{\sigma}$ is any $su(d)$ basis. The result will in fact be independent on the concrete choice of $\obsset{\sigma}$.
It is easy to see that such a covariance matrix of an isotropic state $\Gamma_{\varrho_{\rm iso}}$ contains diagonal blocks
\be
\begin{aligned}
    \gamma_a &= \diag \left( 0 , 1/d , \dots , 1/d \right) = \gamma_b , \\
    X_{\varrho_{\rm iso}} &= \diag \left( 0 , \alpha/d, \dots , \alpha/d \right) .
\end{aligned}
\ee
This can be seen simply from the fact that the isotropic state as written in \cref{eq:isostatesgen} is already in a form analogous to the filtered normal form and we have simply
\be
\begin{aligned}
    \tr\left( \sigma_k\otimes \id \ \varrho_{\rm is} \right) &= \tr\left( \id \otimes \sigma^*_k \ \varrho_{\rm is} \right) = 0 , \\
    \tr\left( \sigma_k \sigma_l\otimes \id \ \varrho_{\rm is} \right) &= \tr\left( \id \otimes \sigma^*_k \sigma^*_l \ \varrho_{\rm is} \right) = \frac{1}{d} \delta_{kl}, \\
    \tr\left( \sigma_k \otimes \sigma^*_l \ \varrho_{\rm is} \right) &= \frac{\alpha}{d} \delta_{kl} . 
\end{aligned}
\ee
Plugging these values into \cref{eq:cor1} we get the condition
\be
\alpha (d^2-1) \leq rd-1 \implies \tr(\varrho_{\rm is} \kb{\psi_+}{\psi_+}) \leq r /d ,
\ee
which is nothing but the expression for the criterion based on the fidelity with respect to the maximally entangled state. 
Thus, for such states we get in an even easier way that the criterion \cref{eq:fidcrit} is just a corollary of the CMC. 

\subsubsection{Rotationally-invariant states}\label{app:rotinvstates}

Let us now consider states that are invariant under unitaries of the type $R_j\otimes R_j$ where $R_j$ is a rotation in the spin-$j$ representation with $j=(d-1)/2$ 
(here again we are considering for simplicity two equal-dimensional quantum systems, but the theory works in the more general setting as well).
From the general theory of invariant states~\cite{VollbrechtWerner01,Schliemann03,Schliemann05,KitranCaves08} we know that they have the form
\be
\varrho_{RR} = \sum_{J=0}^{2j} \frac{\alpha_J}{2J+1} \sum_{M_z=-J}^J \ketbra{J,M_z} ,
\ee
where $\alpha_J\geq 0$ and $\sum_J \alpha_J=1$. For these states it is natural to consider the spin covariance matrix, and it is also easy to calculate. 
Because of the symmetry, we have that
\be
\begin{aligned}
    \tr(j_x \otimes \id \varrho_{RR}) &= \tr(j_y \otimes \id \varrho_{RR}) = \tr(j_z \otimes \id \varrho_{RR})= 0 \\
    \tr(j_x^2 \otimes \id \varrho_{RR}) &= \tr(j_y^2 \otimes \id \varrho_{RR}) = \tr(j_z^2 \otimes \id \varrho_{RR}) = j(j+1)/3 \\
   \tr(j_x \otimes j_x \varrho_{RR}) &= \tr(j_y \otimes j_y \varrho_{RR}) = \tr(j_z \otimes j_z \varrho_{RR}) = \frac 1 6  \left( \sum_{J=0}^{2j} \alpha_J J(J+1) - 2j(j+1)\right) .
\end{aligned}
\ee
The last condition is simply derived from 
\be
\tr\left( (J_x^2 + J_y^2 + J_z^2) \varrho_{RR} \right) = \sum_{J=0}^{2j} \alpha_J J(J+1),
\ee
where $J_k=j_k \otimes \mathbb{1} + \mathbb{1} \otimes j_k,k=x,y,z$. Moreover, these states are also invariant under exchanging the two parties, so we have $\Cov(\obsset{j}_a) = \Cov(\obsset{j}_b)$.

\subsubsection{$4$-partite states invariant under $U\otimes V \otimes U^* \otimes V^*$}\label{app:4partiUUVV}

As a more complex application, that is also relevant to our discussion, let us now consider states in a $4$-particle space $(a_1,b_1,a_2,b_2)$ of dimensions $d_{a_1} = d_{b_1}=d_1$ and $d_{a_2}=d_{b_2}=d_2$, that are invariant under unitaries of the type $U\otimes V \otimes U^* \otimes V^*$. Notable example of such states are the high-Schmidt-number PPT states given in \cref{eq:PPThighSN}.
States with such a symmetry can be expanded as follows
\be\label{eq:UVUstarVstarstategen}
\varrho_{UVU^*V^*} = \delta \id_{d_1}^{\otimes 2}\otimes \id_{d_2}^{\otimes 2} + \alpha_1 \ketbra{\psi_+}_{a_1 b_1} \otimes \id^{\otimes 2}_{d_2} + \alpha_2 \id^{\otimes 2}_{d_1} \otimes \ketbra{\psi_+}_{a_2 b_2} + \beta \ketbra{\psi_+}_{a_1 b_1} \otimes \ketbra{\psi_+}_{a_2 b_2} ,
\ee
where note the $U \otimes U^*$ symmetry applies to the parties $(a_1 b_1)$ and the $V \otimes V^*$ symmetry applies to the parties $(a_2 b_2)$. 

Here, following the isotropic states example, one could consider local basis for each party as follows
\be\label{eq:4PartiteLocalBasis}
\begin{aligned}
    \vec g_{a_1 a_2} &= \left(\frac{\id \otimes \id}{d_a} , \vec{\sigma_a \otimes \id} , \vec{\id \otimes \sigma_a^*}, \vec{\sigma_a \otimes \sigma_a^*} \right) \\
        \vec g_{b_1 b_2} &= \left(\frac{\id \otimes \id}{d_b} , \vec{\sigma_b \otimes \id} , \vec{\id \otimes \sigma_b^*}, \vec{\sigma_b \otimes \sigma_b^*} \right) ,
\end{aligned}
\ee
where $\obsset{\sigma}_a$ and $\obsset{\sigma}_b$ are $su(d_a)$ and $su(d_b)$ bases respectively (which again we don't even need to specify since the result will be independent on the concrete bases).

From the expression \eqref{eq:UVUstarVstarstategen} it is easy to see that the reduced density matrices for the two parts are:
\be
\begin{aligned}
    (\varrho_{UVU^*V^*})_{a_1 a_2} &= \frac 1 {d_1 d_2} \id_{d_1} \otimes \id_{d_2}  = (\varrho_{UVU^*V^*})_{b_1 b_2} ,
\end{aligned}
\ee
which means that their linear entropies are maximal, i.e., $1- \tr\left((\varrho_{UVU^*V^*})^2_{a_1 a_2}\right) 
= 1- \tr\left((\varrho_{UVU^*V^*})^2_{b_1 b_2}\right) = 1 - \tfrac{1}{d_1 d_2}$.

For such states it is also easy to calculate the covariance matrix blocks for the bipartition $(a_1 a_2| b_1 b_2)$, namely $\Cov(\vec g_{a_1 a_2},\vec g_{b_1 b_2})$, 
and afterward to evaluate our \cref{cor:1}. We provide an example relevant to our discussion in the following.

\subsection{PPT entangled states with high Schmidt number}\label{app:PPTstatesSchmidt}

Let us consider the following state
\be\label{eq:PPThighSN}
\varrho^*_{a_1 b_1 a_2 b_2} = d \left( \id - \kb{\psi_+}{\psi_+} \right)_{a_1 b_1} \otimes \left( \id - \kb{\psi_+}{\psi_+} \right)_{a_2 b_2} + d \left( \frac d 2 + 1 \right) \kb{\psi_+}{\psi_+}_{a_1 b_1} \otimes \kb{\psi_+}{\psi_+}_{a_2 b_2} ,
\ee
which is of Schmidt number $s\geq \lceil d/4 \rceil$ across the bipartition $(a_1 a_2 | b_1 b_2)$~\cite{HuberLamiLancienMuellerHermes2018}. 
Note that such a state has a symmetry under unitaries of the type $U\otimes V \otimes U^* \otimes V^*$, and is therefore of the type described earlier in \cref{app:4partiUUVV}. Note also that a Schmidt-number witness that detects this state is given by~\cite{HuberLamiLancienMuellerHermes2018}
\be
W = \ketbra{\psi_+}_{a_1 b_1} \otimes \left( \id - \frac{d}{2r} \ketbra{\psi_+}\right)_{a_2 b_2} .
\ee

Now, one might ask the question whether our corollary \ref{cor:1} would also detect the Schmidt rank of such a state.
In order to check this, let us rewrite the 4-partite PPT state with high Schmidt number \eqref{eq:PPThighSN} in local basis \eqref{eq:4PartiteLocalBasis}.
We have
\begin{equation}
\begin{aligned}
\varrho_{a_1 a_2 b_1 b_2} & =\frac{1}{d^2} \mathbb{1}_{d^2}+\frac{d-4}{d\left(3 d^2+2 d-8\right)} \sum_{i=1}^{\left(\frac{d}{2}\right)^2-1} \mathbb{1}_2 \otimes \sigma_i^{a_2} \otimes \mathbb{1}_2 \otimes\left(\sigma_i^{b_2}\right)^* \\
& +\frac{2\left(-d^2+2 d+8\right)}{d^2\left(3 d^2+2 d-8\right)} \sum_{i=1}^{2^2-1} \sigma_i^{a_1} \otimes \mathbb{1}_{d / 2} \otimes\left(\sigma_i^{b_1}\right)^* \otimes \mathbb{1}_{d / 2} \\
& +\frac{2(d+4)}{d\left(3 d^2+2 d-8\right)} \sum_{i=1}^{2^2-1} \sum_{j=1}^{\left(\frac{d}{2}\right)^2-1} \sigma_i^{a_1} \otimes \sigma_j^{a_2} \otimes\left(\sigma_i^{b_1}\right)^* \otimes\left(\sigma_j^{b_2}\right)^* .
\end{aligned}
\end{equation}
The submatrices of its CM are 
\begin{equation}
\begin{aligned}
\gamma_a & =\operatorname{diag}(0,1 / d, \ldots, 1 / d)=\gamma_b, \\
X_{\varrho_{a_1 a_2 b_1 b_2}}&=\operatorname{diag}(0, \underbrace{\frac{-d^2+2 d+8}{d\left(3 d^2+2 d-8\right)}, \cdots}_3, \underbrace{\frac{2(d-4)}{d\left(3 d^2+2 d-8\right)}, \cdots}_{\frac{d^2}{4}-1}, \underbrace{\frac{2(d+4)}{d\left(3 d^2+2 d-8\right)}, \cdots}_{3\left(\frac{d^2}{4}-1\right)} ).
\end{aligned}
\end{equation}
Let us consider a value of $d\geq 4$. In that case we have that $-d^2+2 d+8 < 0$ and therefore the singular value of the matrix $X$ is given by its opposite, namely $d^2- 2d-8$. Summing up all the singular values of $X$ we get
\be
\frac{\left[3(d^2- 2d-8) + 2(d-4)\left(\frac{d^2}{4}-1\right) + 6(d+4)\left(\frac{d^2}{4}-1\right) \right]}{d\left(3 d^2+2 d-8\right)} = 
\frac{\left[ (d+2)(2d-5)(d+4)\right]}{d\left(3 d^2+2 d-8\right)} .
\ee
Therefore, the left-hand side of \cref{eq:cor1} becomes
\be
\frac{\left[ (d+2)(2d-5)(d+4)\right]}{d\left(3 d^2+2 d-8\right)}  - (r-1) ,
\ee
while the right-hand side is simply $( 1- 1/d )$. Thus, our corollary evaluated on this state becomes
\begin{equation}
r \geq \frac{\left[ (d+2)(2d-5)(d+4)\right]}{d\left(3 d^2+2 d-8\right)} + \frac 1 d = \frac{2 d(d+3)-24}{d(3 d-4)}.
\end{equation}
For any even number $d$, the right-hand-side is less than or equal to $1$. This means even entanglement cannot be detected in this case.

\bibliographystyle{quantum}
\bibliography{Bibliography_quantum.bib}

\end{document}